\documentclass[11pt]{article}
\usepackage[T1]{fontenc}
\usepackage{graphicx} 

\usepackage[utf8x]{inputenc}
\usepackage{amssymb,amsmath,mathrsfs,enumerate,mathtools}
\usepackage{graphicx,rotate,multicol,braket,multirow}
\usepackage{cite}
\usepackage{braket}
\usepackage{array}
\usepackage{url}
\usepackage{comment}
\usepackage{float}
\usepackage{slashed}
\usepackage[normalem]{ulem}
\usepackage{wrapfig}
\usepackage[textfont=it,labelfont=bf]{caption}
\usepackage{bm}
\usepackage{cancel}
\usepackage[colorlinks=true,
linkcolor=blue,
urlcolor=blue,
citecolor=teal]{hyperref}
\usepackage{lineno}
\usepackage{color}
\usepackage{bbold}
\usepackage{orcidlink}           

\long\def\rpl#1!!#2!!{\textcolor{red}{#1} \textcolor{blue}{#2}}

\def\bar{\overline}
\def\tilde{\widetilde}

\evensidemargin=\oddsidemargin
\title{Investigating the leptonic couplings of doubly charged scalars \\ at the muon collider}
\author{
        \sf
      Nivedita Ghosh$^{a,}$\footnote{\href{mailto:nivedita.ghosh@ipmu.jp}{nivedita.ghosh@ipmu.jp}}\ \orcidlink{0000-0001-8469-3227},
      Santosh Kumar Rai$^{b,}$\footnote{\href{mailto:skrai@hri.res.in}{skrai@hri.res.in}}\ \orcidlink{0000-0002-4130-6992},
      Tousik Samui$^{c,}$\footnote{\href{mailto:tousiks@imsc.res.in}{tousiks@imsc.res.in}}\ \orcidlink{0000-0002-1485-6155},
      Agnivo Sarkar$^{b,}$\footnote{\href{mailto:agnivosarkar@hri.res.in}{agnivosarkar@hri.res.in}}\ \orcidlink{0000-0001-9596-1936}
      \\[3mm]
      \small\em
      $^a$Kavli IPMU (WPI), UTIAS, University of Tokyo, Kashiwa, Chiba, 277-8583, Japan\\
      \small\em
	$^b$Regional Centre for Accelerator-based Particle Physics, Harish-Chandra Research Institute,\\
	\small\em
	 A CI of Homi Bhabha National Institute, 
    Chhatnag Road, Jhunsi, Prayagraj 211019, India\\
         \small\em 
         $^c$The Institute of Mathematical Sciences, IV Cross Road, CIT Campus, Taramani, Chennai 600113, India
        }

\date{}

\begin{document}

\begin{flushright}
	\small{HRI-RECAPP-2025-04}\\
        \small{IMSc/2025/02}
\end{flushright} 

{\let\newpage\relax\maketitle}	

\begin{abstract}
\noindent
We study the lepton flavor conserving and violating couplings of a doubly charged scalar at a 3 {\tt TeV} muon collider. Using a model-independent Lagrangian, we analyze the $e\,e$, $\mu\,\mu$, and $\tau\,\tau$ final states mediated by the doubly charged scalar to probe individual couplings to $\mu\,e$, $\mu\,\mu$, and $\mu\,\tau$. We find that for a doubly charged scalar of mass greater than 1 {\tt TeV} and $\mathcal{O}(1)$ couplings, we achieve high signal significance in these channels. We delineate the collider’s sensitivity in the mass-coupling plane, highlighting the extensive reach of the muon collider in probing these couplings far beyond the current experimental limits. We also propose an angular distribution variable to discriminate between the exchange of a doubly charged scalar from that of a neutral scalar, which gives identical signals. 
\end{abstract}


\section{Introduction}
The discovery of the Higgs boson at the Large Hadron Collider (LHC) \cite{ATLAS:2012yve,CMS:2012qbp} marked a significant milestone in particle physics, confirming the mechanism by which elementary particles acquire mass within the Standard Model (SM). 
However, notwithstanding its various experimental confirmations, SM still leaves many unanswered questions, including the mechanism for neutrino masses.
To address this, various extensions to the SM have been proposed in the literature, one of which is the Higgs Triplet Model (HTM) \cite{Ross:1975fq,Gunion:1989ci,Nomura:2022wck}. This model introduces a triplet scalar field, which includes a doubly charged Higgs boson. The primary motivation for considering the HTM and the doubly charged Higgs lies in its role within the type-II seesaw mechanism \cite{Magg:1980ut}, which provides a natural explanation for the smallness of neutrino masses through the vacuum expectation value (VEV) of the scalar triplet. Beyond the HTM, the doubly charged Higgs also appears in the Left-Right Symmetric Model (LRSM), which extends the gauge group to $SU(2)_L \times SU(2)_R \times U(1)_{B-L}$, proposing a symmetry between left- and right-handed particles and includes triplet scalars \cite{Mohapatra:1974gc,Senjanovic:1975rk,Mohapatra:1979ia,Deshpande:1990ip,Brahmachari:2003wv}. In LRSM, both the $SU(2)_L$ and $SU(2)_R$ scalar triplets can include doubly charged components, with the $SU(2)_R$ triplet's VEV potentially at the TeV scale to provide masses for right-handed gauge bosons. The LRSM model also addresses the strong CP problem and parity violation, further motivating the study of doubly charged Higgs bosons. Other notable models include doubly charged scalars, which can generate neutrino masses through a radiative mechanism \cite{Mohapatra:1987nx,Balakrishna:1988bn,Ma:2006km,Gu:2007ug,Saad:2019bqf,Babu:2019mfe,Aoki:2020eqf,Maharathy:2022gki,Giarnetti:2023dcr,Giarnetti:2023osf,Jana:2024icm,Borboruah:2025bwx}.

The ATLAS \cite{ATLAS:2012hi,ATLAS:2014kca,ATLAS:2017xqs,ATLAS:2018ceg,ATLAS:2021jol} and CMS \cite{CMS:2012dun,CMS:2014mra,CMS:2016cpz,CMS:2017pet,CMS:2017fhs} collaborations have searched for these doubly charged scalars $H^{\pm\pm}$ in both pure leptonic as well as $W^{\pm}W^{\pm}$ final states. One can note that in the majority of these searches, pair production of the doubly charged scalars via the Drell-Yan mode remains the primary production mode. A doubly charged scalar can also be produced at the LHC via associated and vector boson fusion modes. However, in comparison to Drell-Yan mode, these production channels depend on how the electroweak symmetry is broken in the underlying model. A comprehensive analysis considering all these channels at the LHC and their sensitivity in the mass vs.~coupling plane $\{m_{H^{\pm\pm}}, h_{XY}\}$ plane can be found in Ref.~\cite{Ashanujjaman:2021txz}. The experiment has been able to probe the doubly charged scalar up to masses of $\mathcal{O}({\tt TeV})$. The major obstacle that the LHC faces in excluding very large values of $m_{H^{\pm\pm}}$ or much smaller values $h_{XY}$ couplings is due to the center of mass ($\tt{c.o.m}$) being not sufficient enough to pair produce these relatively heavier leptophilic exotic scalars with large enough rates. 
Thus, one has to look for an alternative collider that can surpass these challenges of the pure hadronic machine. 

In contrast, if one considers a leptonic collider\cite{LinearColliderAmericanWorkingGroup:2001rzk,ECFADESYLCPhysicsWorkingGroup:2001igx,LinearColliderACFAWorkingGroup:2001awr,CLICPhysicsWorkingGroup:2004qvu,ILC:2013jhg,Moortgat-Pick:2015lbx,LinearColliderVision:2025hlt,LinearCollider:2025lya}, then one can think of an alternative signal channel {\sl e.g.} $e^{+}_{a}e^{-}_{a} \to e^{+}_{i}e^{-}_{j}$ (where $a = e, \mu$ and $i,j = e, \mu, \tau$), where the doubly charged scalar participates in a $t$-channel process. The advantage of this channel is two fold -- (\emph{i}) \emph{in lieu} of pair production, a single $H^{\pm\pm}$ is participating, and (\emph{ii}) as the signal channel is mediated via $t$-channel $H^{\pm\pm}$, this machine will be able to probe $m_{H^{\pm\pm}} \geq \sqrt{s}$ ($\tt{c.o.m}$ of the lepton collider) region. These aspects motivate us to conduct our study in the context of a lepton collider \cite{Godbole:1994np,Cheung:1994rp,Yagyu:2014aaa,Chiang:2015rva,Blunier:2016peh,Crivellin:2018ahj,Bai:2021ony,Kumar:2024ghy,Altmannshofer:2025nbp}. 
If one considers the first generation lepton collider, that is $e^{+}e^{-}$ circular collider, then at very high energies, the large synchrotron radiation will restrict the availability for very high energy collisions significantly, limiting the reach for BSM searches. Even at the proposed linear colliders for $e^{+}e^{-}$ collisions, beam radiation will be a major issue. A second-generation lepton collider (or, in other words, $\mu^{+}\mu^{-}$ collider) seems like the next best alternative that could achieve very high-energy collisions with much less synchrotron radiation or beam effects.   

The proposal for the high-energy muon collider by the \textit{International Muon Collider Collaboration}~\cite{IMCC,Delahaye:2019omf,Schulte:2019bdl,Schulte:2022brl,InternationalMuonCollider:2024jyv,Begel:2025ldu} is a timely 
proposal going worldwide. The reason behind the popularity of this collider comes from the fact that it can offer the advantages of both the lepton collider as well as the hadron collider~\cite{Costantini:2020stv,Han:2020uid,Han:2021kes,AlAli:2021let,MuonCollider:2022xlm,Accettura:2023ked,Barik:2024kwv,Abe:2025yur}. To elaborate on the advantages, let us first compare it to hadron colliders. The muon collider will ideally be more efficient at high energies because it can use all of the machine's energy, unlike hadron colliders, where the beam energy is shared among the partons that participate in the hard scattering processes with effectively lower energies. Also, at the muon colliders, one expects to have a much more precise kinematic information regarding the initial state colliding particles. In addition, hadron colliders face a lot of unsolicited noise from hadronic activity and the smearing caused by parton distribution functions (PDFs), making precision studies hard. The muon can produce high center-of-mass energies in collisions with very little energy spread due to reduced bremsstrahlung and beamstrahlung effects~\cite{Chen:1993dba,Barklow:2023iav}, as the mass of the muon is appreciably higher than that of the electron. Although the energy and luminosity of the muon collider are not finalized yet, there is proposal to run it with $\tt{1 ~ab^{-1}}$ luminosity at $\tt{3~TeV}$ $\tt{c.o.m}$ energy and $\tt{10~ab^{-1}}$ luminosity for a $\tt{10~TeV}$ machine~\cite{Costantini:2020stv,Han:2020uid,InternationalMuonCollider:2024dlm}. With this first-ever proposal of a second-generation lepton collider, one hopes to discover new physics ($\tt{NP}$) scenarios that HL-LHC might not be able to probe \cite{Braathen:2020frt,Chiesa:2021qpr,Ghosh:2022vpb,Franceschini:2022dhb,Mekala:2023diu,Belyaev:2023yym,Li:2023lkl,Ghosh:2023xbj,Asadi:2023csb,Bi:2024pkk,Bandyopadhyay:2024gyg,Zhao:2024fgc,Han:2025wdy,InternationalMuonCollider:2025sys,Abu-Ajamieh:2025zcv,Yang:2025jxc}.

In this paper, our goal is to investigate the possible physics prospect that the future muon collider offers, specifically targeting the scalar sector of different BSM models, which can embed a leptophilic charged scalar. To do so, we will consider a class of BSM scenarios where the SM particle spectrum is enlarged with at least one pair of doubly charged scalar particles. In the context of LHC, there are already many recent works that search for the doubly charged scalars~\cite{Mitra:2016wpr,Ghosh:2017pxl,KumarGhosh:2018bli,Padhan:2019jlc,Jueid:2023qcf}. However, dedicated analyses to probe the doubly charged scalar and its couplings at a muon collider are very few in the literature \cite{Li:2023ksw,Maharathy:2023dtp,Jia:2024wqi}.

\vspace{0.2cm}

\noindent
In this work, we attempt to find answers to the following questions:
\begin{enumerate}
\item Can a future muon collider with 1 $\rm ab^{-1}$ luminosity at $\tt{3~TeV}$ $\tt{c.o.m}$ energy show significant improvement in terms of probing the leptonic coupling of the doubly charged scalar? 

\item Since we focus on an off-shell exchange of the doubly charged scalar, there could potentially be other $\tt{(NP)}$ particles that contribute to similar final states that we choose for our analysis. What are the possible ways to disentangle the $\tt{NP}$ source in the context of the current study at the muon collider?

\end{enumerate}

The article is organized in the following fashion - in section \ref{sec:model} we introduce the theoretical set up for our study and determine the allowed region in the $\{m_{H^{\pm\pm}}, h_{XY}\}$ parameter plane after imposing the existing flavor bounds as well as different direct search limits. After pinning down the allowed region of the parameter space, we present the detailed analysis in section \ref{sec:analysis} for the $\tt{3~TeV}$ muon collider. For present discussion, we focus on three signal events $\mu^{+} \mu^{-} \to \mu^{+} \mu^{-}$, $\mu^{+} \mu^{-} \to e^{+} e^{-}$ and $\mu^{+} \mu^{-} \to \tau^{+} \tau^{-}$ as these channels enable us to impose absolute bounds on $\{h_{\mu\mu}, h_{\mu e}, h_{\mu\tau}\}$ couplings, respectively, \emph{w.r.t.} doubly charged scalar mass. In section \ref{sec:summ}, we illustrate the corresponding reach plots in the aforementioned parameter plane. Before concluding our discussion, in section \ref{sec:inverse} we demonstrate a crucial aspect of this muon collider in terms of distinguishing two different $\tt{NP}$ scenarios. Finally, we summarize our study in section \ref{sec:conc}.


\section{Setup}
\label{sec:model}
\begin{figure}[ht!]
\centering
\includegraphics[height=5.0cm,width=8.0cm]{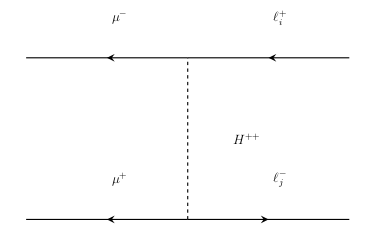}
\caption{The $t$-channel Feynman diagram correspond to the $H^{\pm\pm}$ mediated signal process $\mu^{+} \mu^{-} \to \ell^{+}_{i} \ell^{-}_{j}$, where the index stands for $i, j = e, \mu, \tau$.}
\label{Fig:feynman}
\end{figure}

\noindent 
The interaction Lagrangian, which is relevant for our study, will take the following form:
\begin{equation}
\mathcal{L} \supset \sum_{i,j = e,\mu, \tau}h_{ij}\ell_{iL}^{T}\mathcal{C}H^{++}\ell_{jL} + \text{h.c}.
\label{Eqn:theory}
\end{equation}
Here, the coefficient $h_{ij}$ indicates leptonic couplings of the doubly charged scalar, and $\mathcal{C}$ denotes the charge conjugation operator.  Without loss of generality, $h_{ij}$ is taken to be symmetric in the flavor indices $i$ and $j$, and the couplings are further assumed to be real.
This \emph{phenomenological} choice of the Lagrangian term can be realized in various scalar extended BSM scenarios, e.g., hybrid seesaw~\cite{Schechter:1980gr,Dev:2019hev}, Zee-Babu~\cite{Zee:1985id,Babu:1988ki}, and left-right symmetric model \cite{Mohapatra:1974gc,Senjanovic:1975rk,Mohapatra:1979ia,Deshpande:1990ip,Brahmachari:2003wv,Maharathy:2022gki,Borboruah:2025bwx} to name a few. While certain models involve right-chiral leptons as opposed to the left-chiral structure assumed here, the resulting collider signatures considered in this work under unpolarized beams remain effectively the same, and our analysis does not depend on the specific chiral assignment. Depending on the model, one can set the explicit value of $h_{ij}$ and the doubly charged scalar mass $m_{H^{\pm\pm}}$. Instead of focusing on any specific model, we attempt to probe the above interaction maximally using the proposed muon collider. We find that the muon machine helps in imposing an absolute bound on the values of $h_{\mu\tau}, h_{\mu\mu}$ couplings, which are still missing in the literature. To emphasize this point, we present in Table~\ref{Tab:flavour}, the different flavor-violating processes that could be influenced by the presence of a doubly charged scalar that couples to the charged leptons. The table includes constraints on the scalar’s couplings derived from these processes~\cite{BhupalDev:2018tox,Cheng:2022jyi}. It is worth noting that the flavor-changing decays such as $\ell_{i} \to \ell_{j}\ell_{k}\bar{\ell}_{l}$, where $\ell$ represent the charged leptons only, are absent in the SM. 
\begin{table}[ht!]
\centering
\begin{tabular}{| m{7em} || c | c | c |}
\hline
Process & Experimental Bound & Constraint on & Bound $\times\left(m_{H^{\pm\pm}}/\text{TeV}\right)^{2}$ \\
\hline \hline
 $\mu \to e\gamma$ & $< 4.2\times10^{-13}$ \cite{MEG:2016leq} & $\left|\sum_{k = e, \mu, \tau}h^{\dagger}_{k\mu}h_{k e}\right|$ & $< 2.4\times10^{-4}$ \\ 
\hline
$\mu \to 3e$ & $< 1.0\times10^{-12}$ \cite{SINDRUM:1987nra} & $\left|h_{\mu e}h_{ee}\right|$ & $< 2.3\times10^{-5}$  \\
\hline
$\tau \to e\gamma$ & $< 3.3\times10^{-8}$ \cite{BaBar:2009hkt} & $\left|\sum_{k = e, \mu, \tau}h^{\dagger}_{ke}h_{k \tau}\right|$ & $< 1.6\times10^{-1}$  \\
\hline
$\tau \to \mu\gamma$ & $< 4.2\times10^{-8}$ \cite{Belle:2021ysv} & $\left|\sum_{k = e, \mu, \tau}h^{\dagger}_{k\mu}h_{k \tau}\right|$ & $< 1.9\times10^{-1}$  \\
\hline
$\tau \to 3e$ & $< 2.7\times10^{-8}$ \cite{Hayasaka:2010np} & $\left|h_{\tau e}h_{ee}\right|$ & $< 9.2\times10^{-3}$  \\
\hline
$\tau \to \mu^{+}\mu^{-}e^{-}$ & $< 2.7\times10^{-8}$  \cite{Hayasaka:2010np} & $\left|h_{\tau \mu}h_{\mu e}\right|$ & $< 6.5\times10^{-3}$  \\
\hline
$\tau \to e^{+}\mu^{-}\mu^{-}$ & $< 1.7\times10^{-8}$  \cite{Hayasaka:2010np} & $\left|h_{\tau e}h_{\mu \mu}\right|$ & $< 7.3\times10^{-3}$  \\
\hline
$\tau \to e^{+}e^{-}\mu^{-}$ & $< 1.8\times10^{-8}$  \cite{Hayasaka:2010np} & $\left|h_{\tau e}h_{\mu e}\right|$ & $< 5.3\times10^{-3}$  \\
\hline
$\tau \to \mu^{+}e^{-}e^{-}$ & $< 1.5\times10^{-8}$  \cite{Hayasaka:2010np} & $\left|h_{\tau\mu}h_{ee}\right|$ & $< 6.9\times10^{-3}$  \\
\hline
$\tau \to 3\mu$ & $< 2.1\times10^{-8}$  \cite{Hayasaka:2010np} & $\left|h_{\tau\mu}h_{\mu\mu}\right|$ & $< 8.1\times10^{-3}$  \\
\hline\hline
$\text{M} \to \overline{\text{M}}$ & $\leq 8.2\times10^{-11}$ \cite{Willmann:1998gd} & $\left|h_{ee}h_{\mu\mu}\right|$ & $< 4.9\times10^{-2}$ \\
\hline \hline 
$e e \to e e$ [LEP] & $\Lambda_{\tt eff} > 5.2$ TeV \cite{DELPHI:2005wxt} &  $\left|h_{ee}\right|^{2}$ & $< 1.2\times10^{-1}$  \\
\hline
$e e \to \mu \mu$ [LEP] & $\Lambda_{\tt eff} > 7.2$ TeV \cite{DELPHI:2005wxt} &  $\left|h_{\mu e}\right|^{2}$ & $< 6.4\times10^{-2}$  \\
\hline
$e e \to \tau \tau$ [LEP] & $\Lambda_{\tt eff} > 7.6$ TeV \cite{DELPHI:2005wxt} &  $\left|h_{e \tau}\right|^{2}$ & $< 5.4\times10^{-2}$  \\
\hline
\end{tabular}
\caption{The updated bounds on different flavor-violating as well as flavor-conserving processes, which were measured by different experimental collaborations. From the last column, one can notice that all these bounds depend on the underlying doubly charged scalar mass.}
\label{Tab:flavour}
\end{table}
\noindent 
In contrast, these modes can emerge even at the tree-level in models with doubly charged scalars that allow flavor-violation in the leptonic decay modes. In principle, neutral scalars can also induce such FCNC processes, for instance in MSSM-type SUSY models~\cite{Kersten:2014xaa} and in several variants of the Two-Higgs-Doublet Model (2HDM)~\cite{Davidson:2010xv,Davidson:2016utf}. In this work, however, we focus mainly on the phenomenology of the doubly charged scalar. On the other hand, for processes like $\ell_{i} \to \ell_{j}\gamma$, that can arise within the SM at 1-loop via the 
$W$-boson exchange, one gets additional contributions from the doubly charged scalar that can modify the corresponding 
cross-section\footnote{There can be an additional contribution from singly charged scalars, which is approximately 8 times smaller than that of the doubly charged scalar~\cite{Akeroyd:2009nu,Lindner:2016bgg}. Thus, the dominant contribution comes from the doubly charged scalars in these processes.}. Apart from these lepton flavor-violating (LFV) processes, the Large Electron-Positron ($\tt{LEP}$) collider facility also searched for doubly charged scalars in the $e^{+}e^{-} \to \ell^{+}_{i}\ell^{-}_{i}$ channels. It is important to note that the majority of these above-mentioned processes are primarily proportional to the multiplicative coupling factor $h_{ij}h^{*}_{kl}$, where each index represents the lepton families. As mentioned before, the muon collider provides us with the scope to 
probe to significant sensitivity, the individual $\{h_{\mu\tau}, h_{\mu\mu}, h_{\mu e}\}$
couplings that can be realized from the Feynman diagram illustrated in Figure \ref{Fig:feynman}, 
very similar to 
$\tt{LEP}$ which imposed absolute bounds on $\{h_{ee}, h_{e\mu}, h_{e\tau}\}$.   
One can always choose a fine-tuned value for the $h_{ij}$ (where $i,j = e, \tau$) to satisfy these constraints while keeping the value of $\{h_{\mu\tau}, h_{\mu\mu}, h_{\mu e}\}$ at $\mathcal{O}(1)$. To emphasize this point, let us consider the muonium-antimuonium transition $M \to \overline{M}$~\cite{Willmann:1998gd}. Here, the upper bound on the probability of conversion between the LFV bound states (\emph{i.e.} $\mu^{+}e^{-} \longleftrightarrow \mu^{-}e^{+}$) will take the following form 
\begin{equation}
\frac{\left|h_{ee}h_{\mu\mu}\right|}{\left(m_{H^{\pm\pm}}/\text{TeV}\right)^{2}} \leq 4.9\times10^{-2}.
\label{Eq:MMbar}
\end{equation}
If we consider a BSM scenario which contains a doubly charged scalar of mass $m_{H^{\pm\pm}} = 1~\tt{TeV}$ and choose $h_{ee} = \mathcal{O}\left(10^{-3}\right)$, then the corresponding value of $h_{\mu\mu}$ can be fixed at $\mathcal{O}\left(1\right)$ while respecting the relevant limit. Similarly, the $e e \to e e$ bound from LEP suggests that $\left|h_{ee}\right| < 0.346$ for the $m_{H^{\pm\pm}} = 1~\tt{TeV}$.

\begin{figure}[ht!]
\centering
\includegraphics[scale=0.38]{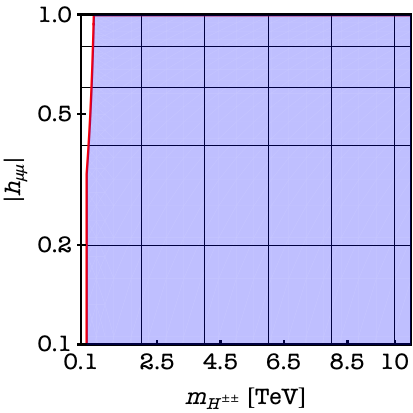}
\includegraphics[scale=0.38]{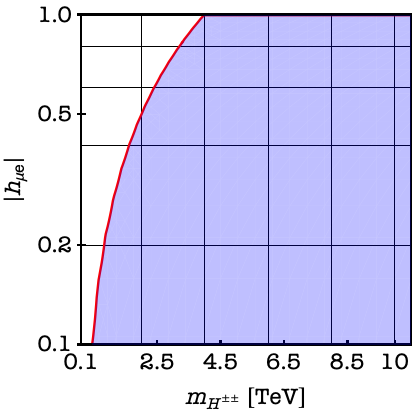}
\includegraphics[scale=0.38]{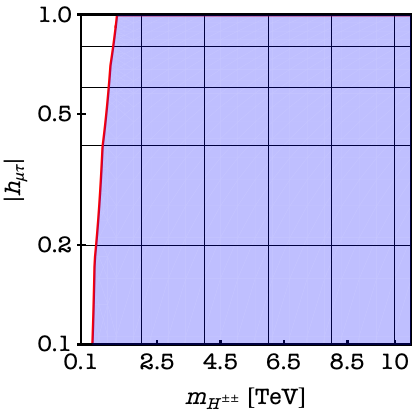}
\caption{The allowed region in the $m_{H^{\pm\pm}}~vs.~h_{\mu i}$ (where $i = \mu, e, \tau$) plane after simultaneously satisfying all relevant flavor data shown in Table \ref{Tab:flavour}. The blue shaded region represents the allowed region of the parameter space after imposing available flavor bounds.}
\label{Fig:flavour}
\end{figure}

Using the given bounds in Table \ref{Tab:flavour}, we determine the excluded region in the parameter plane of $\{m_{H^{\pm\pm}},\,  h_{\mu i}\}$ (where $i = \mu, e, \tau$).
In Figure \ref{Fig:flavour} we present the maximum allowed value\footnote{To perform the scan on $h_{\mu i}$ we have fixed the upper bound $h_{\mu i} < 8\pi$. This choice is motivated from the perturbativity constraint \cite{Branco:2011iw}.} of $h_{\mu i}$ (where $i = \mu, e, \tau$) with respect to doubly charged scalar mass $m_{H^{\pm\pm}}$. As expected, both the $h_{\mu\mu}$ and $h_{\mu \tau}$ can be $\mathcal{O}(1)$ without violating any low-energy experimental limits. In contrast, $h_{\mu e}$ has the strongest limit and is constrained to values less than $\mathcal{O}\left(1\right)$ for  $m_{H^{\pm\pm}} \lesssim 4~\tt{TeV}$. This is expected as $ee \to \mu\mu$ scattering observed at $\tt{LEP}$ imposes an absolute bound on the coupling $h_{\mu e}$ while no such bound is available for the other two couplings.

We also note that there exist projections~\cite{Asadi:2025dii} for future proposed flavor experiments that can constrain some of the quantities listed in Table~\ref{Tab:flavour}. For some observables, these projected limits are significantly stronger than the current bounds. However, we have explicitly checked that these improvements do not affect the bounds in the $h_{\mu i}$-$m_{H^{\pm\pm}}$ plane shown in Figure~\ref{Fig:flavour}. This is because these projections typically constrain combinations of couplings, and such bounds can be satisfied by appropriately readjusting one coupling without violating the others. For example, one of the strongest constraints arises from the LFV process $\mu \to 3e$, for which the projected sensitivity is of order $\mathcal{O}(10^{-16})$~\cite{Mu3e:2020gyw}. This leads to an upper bound $|h_{\mu e} h_{ee}| \lesssim 10^{-7} \times m_{H^{\pm\pm}}$. Since this constraint applies to the product of couplings, one can always choose a sufficiently small value of $h_{ee}$ to satisfy the projected limit without altering the corresponding value of $h_{\mu e}$.
	
Furthermore, flavor experiments predominantly probe the low $m_{H^{\pm\pm}}$ region and are largely insensitive to the individual coupling strengths (see Figure~\ref{Fig:flavour}). In contrast, collider experiments have the advantage of probing the high mass and high coupling regions. In this context, we emphasize that a muon collider provides a unique opportunity to explore complementary regions of parameter space, particularly at large masses and large couplings. It can also place bounds on individual couplings, similar to those obtained at LEP~\cite{DELPHI:2005wxt}, rather than on combinations of couplings as in flavor experiments. Moreover, since the muon collider involves second-generation leptons in the initial state, it is especially well suited to probe muon-related couplings, allowing the $h_{\mu i}$ couplings to be tested effectively.

Before concluding this section, we would like to point out that both ATLAS and CMS Collaborations have searched for these doubly charged scalars \cite{CMS:2017pet,ATLAS:2022pbd} in the pure leptonic decay modes and found no significant excess after analyzing the $\tt{Run~2}$ data.
The null result of these analyses imposes a lower limit on $m_{H^{\pm\pm}}$, which is roughly around $1065~\tt{GeV}$ from ATLAS in the LRSM case, irrespective of the leptonic final state. The analysis uses the pair production of the doubly charged scalar mediated by the neutral gauge bosons in the model. 
These limits are model dependent, and for example, in the case of the Zee-Babu model \cite{Zee:1985id}, the lower limit is relaxed to $880~\tt{GeV}$ due to the absence of the $Z$-mediated diagram. 


\section{Analysis}\label{sec:analysis}
In this section, we provide a detailed analysis of the potential of a muon collider operating at $\sqrt{s} = 3~\tt{TeV}$ to thoroughly explore the $\{m_{H^{\pm\pm}}, h_{\mu i}\}$ parameter space.
To do so, we consider three different signal topologies \emph{i.e.} $\mu^{+}\mu^{-} \to e^{+}e^{-}$, $\mu^{+} \mu^{-} \to \mu^{+}\mu^{-}$ and $\mu^{+} \mu^{-} \to \tau^{+} \tau^{-}$ as illustrated in the Feynman diagram of Figure \ref{Fig:feynman}. The irreducible as well as reducible SM processes, which can make the search process challenging, are listed in the Table \ref{Tab:SMBkg} along with their corresponding production rates at the muon collider. To analyze the strength of the cross-sections shown in Table \ref{Tab:SMBkg}, it is essential to examine the underlying subprocesses contributing to each. 
The $\tau\tau$ background arises primarily via an $s$-channel process, with its cross-section suppressed by the propagator ($\gamma$ and $Z$) as 
$\sqrt{s} = 3~\tt{TeV} \gg M_{Z, \, \gamma} $. The $Z\ell\ell$ and $W\ell\nu$ backgrounds, on the other hand, give reduced cross-sections due to a combination of phase space suppression and the involvement of additional $\alpha_{EW}$ or $\alpha_{em}$ in the $2 \to 3$ process. In contrast, the $\ell^{+}\ell^{-}$ can be generated through both $s$-channel (which is again suppressed due to reasons mentioned above) and $t$-channel diagrams. The $t$-channel contributions to the cross-section are significant for small scattering angles and constitute the most substantial background in our analysis.    
\begin{table}[t!]
\centering
\begin{tabular}{||c | c | c | c ||}
\hline
~~~\tt{$\tau^{+}\tau^{-}$}~~~ & ~~~\tt{$Z\ell^{+} \ell^{-}$}~~~ & ~~\tt{$W\ell\nu $}~~ & ~~~\tt{$\ell^{+} \ell^{-}$}~~~ \\ [0.5ex]
\hline \hline
\tt{11.71 fb} & \tt{57.17 fb} & \tt{1.909 pb} & \tt{13.25 pb} \\
\hline
\end{tabular}
\caption{SM background cross-section at the muon collider with the \emph{c.o.m} energy $\sqrt{s} = 3$ TeV. 
}
\label{Tab:SMBkg}
\end{table}
To calculate the signal cross-section, one can use the explicit formula~\cite{Babu:1988ki} (neglecting the muon mass),
\begin{equation}
\left(\frac{d\sigma}{d\Omega}\right)_{\text{Fig.}(\ref{Fig:feynman})} = \left(\frac{h^{2}_{\mu i}}{4\pi}\right)^{2}\frac{1}{s}\left[\frac{s^{2}\left(1 + \cos\theta\right)^{2}}{\left(s + s\cos\theta + 2m^{2}_{H^{\pm\pm}}\right)^{2} + m^{2}_{H^{\pm\pm}} \Gamma^{2}_{H^{\pm\pm}}}\right].
\label{Eq:signalxsec}
\end{equation}
Here, the label $i$ can be varied depending on the final state leptons, and $\Gamma_{H^{\pm\pm}}$ is the total decay width of the scalar $H^{\pm\pm}$. The $\theta$ is angle between initial state $\mu^{-}$ and $\ell^{-}_{i}$. To generate signal processes, we incorporate the interaction Lagrangian of Eq.(\ref{Eqn:theory}) in $\tt{FEYNRULESv2.3.x}$\cite{Alloul:2013bka} and create the necessary $\tt{UFO}$ file. To estimate the cross-section for the signal and the SM backgrounds, we use $\tt{MadGraph5@NLOv2.9.21}$\cite{Alwall:2011uj}.

\begin{figure}[h!]
\centering
\includegraphics[scale=0.4]{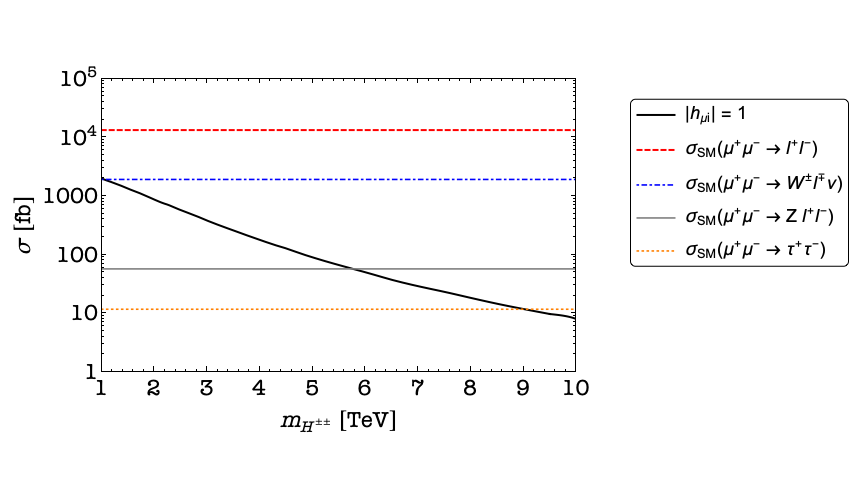}
\caption{Variation of cross-section at the 3 TeV muon collider corresponding to $t$-channel $H^{\pm\pm}$ exchange diagram (Fig.~\ref{Fig:feynman}) as a function of $m_{H^\pm\pm}$.
We have set $h_{\mu i} =1$, and the index $i$ in $\left|h_{\mu i}\right|$ represents $i = e, \mu, \tau$, respectively. The horizontal lines correspond to cross-sections of various SM background processes.}
\label{Fig:sigxsec}
\end{figure}
In Figure \ref{Fig:sigxsec} we present the cross-sections for the doubly charged scalar-mediated process by the black solid line as a function of the scalar mass. The SM background contributions are independent of the scalar mass and their values are constant for the given c.o.m. energy and are shown by the horizontal lines {\sl viz.} $\ell^{+}\ell^{-}$ (red dashed line), $W^{\pm}\ell^{\mp}\nu$ (blue dot-dashed), $Z\ell^{+}\ell^{-}$ (gray solid line) and $\tau^{+}\tau^{-}$ (orange dotted line), respectively. The steady fall of the signal cross-section can be understood from Eq.(\ref{Eq:signalxsec}) where the corresponding $\sigma$ scales approximately as $\sim \left(\frac{1}{m^{4}_{H^{\pm\pm}}}\right)$. Although the $\ell^{+}\ell^{-}$ background appears dominant, we note that the background contributes significantly only when the leptons do not change flavor. So at a muon collider, the large $t$-channel contribution exists exclusively for the $\mu^+\,\mu^-$ final state. In all other scenarios involving lepton flavor change between the initial and final states, it remains insignificant. For a brief comparison, we also comment on the pair production of the doubly-charged Higgs in Drell-Yan mode, a commonly studied search channel at colliders. For a mass of 1250 and 1450 GeV, the cross-sections are 2.2 fb and 0.2 fb, respectively. 
To generate events for both signal and SM background using $\tt{MadGraph5@NLO}$, we impose the following pre-selection cuts 
\begin{equation}
\tt{\left|\eta(\ell)\right| < 3.0;~~~ p_{T}(\ell) > 10~GeV;~~~ \Delta R_{\ell\ell} \geq 0.2}
\label{Eq:preselection}
\end{equation}
After simulating the events, we pass them to $\tt{PYTHIA8}$\cite{Sjostrand:2014zea}, which helps include final state radiations (FSR) and generates event shapes and distributions of the final state events to be used in detector simulations. The detector simulation is performed using $\tt{DELPHESv3.5.0}$\cite{deFavereau:2013fsa} with the help of the appropriate muon collider card~\cite{muoncard}.  
For each of the signals, we try to identify suitable observables that would warrant us to devise a suitable cut, which in turn would help us to reject SM backgrounds maximally while affecting the signal numbers minimally. Finally, we use Eq.(\ref{Eq:significance}) to evaluate the significance of our proposed cut-flow chart~\cite{Cowan:2010js}.       
\begin{equation}
\mathcal{Z} = \left(2\mathcal{L}\right)^{\frac{1}{2}}\left[(\sigma_{S} + \sigma_{B})\text{ln}\left(1 + \frac{\sigma_{S}}{\sigma_{B}}\right) - \sigma_{S}\right]^{\frac{1}{2}}
\label{Eq:significance}
\end{equation}
Here $\mathcal{Z}$ represents significance and $\mathcal{L}$ is the integrated luminosity of the collider. The $\sigma_{S}$ and $\sigma_{B}$ represent the effective cross-section of the signal and background, respectively, after implementing all the cuts. In the following, we will present a detailed description of each final state separately.

We further note that the $e^+ e^-$, $\mu^+ \mu^-$, and $\tau^+\tau^-$ production cross-sections receive additional contributions from the interference with the SM $\gamma/Z$ exchange diagrams. The following analyses presented in this section and the next section include these interference effects. A detailed discussion of interference effects and the relative changes with respect to the $H^{\pm\pm}$-exchange-only contribution is provided in Appendix~\ref{sec:appA}.

\subsection{\texorpdfstring{$\mu^{+}\mu^{-}$}{mu+ mu-} channel}
We begin our discussion with the $\mu^{+}\mu^{-}$ signal topology while selecting exactly two muons in the final state. We impose the following kinematic selection criteria on the final state muons:
\begin{equation}
\tt{p_{T}\left(\mu\right) > 50~GeV; ~~~ |\eta\left(\mu\right)| < 2.4}
\label{Eq:muconstract}
\end{equation}
The cross-section for the signal arising from the doubly charged scalar exchange is calculated with the scalar mass as $m_{H^{\pm\pm}} = 1250~\tt{GeV}$ and assuming the Yukawa coupling to be $h_{\mu\mu} = 1$. 
\begin{figure}[t!]
\centering
\includegraphics[scale=0.24]{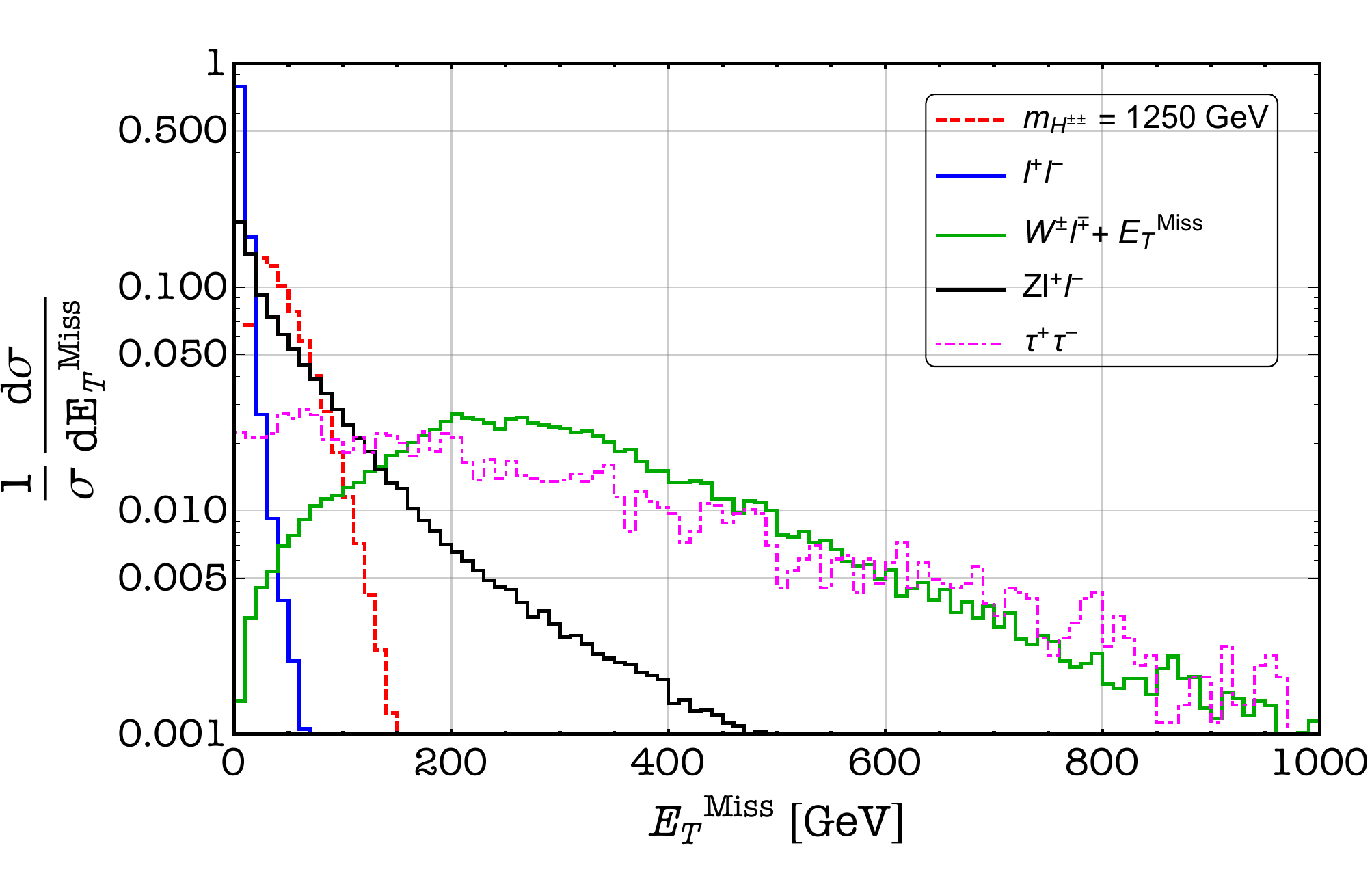}
\includegraphics[scale=0.24]{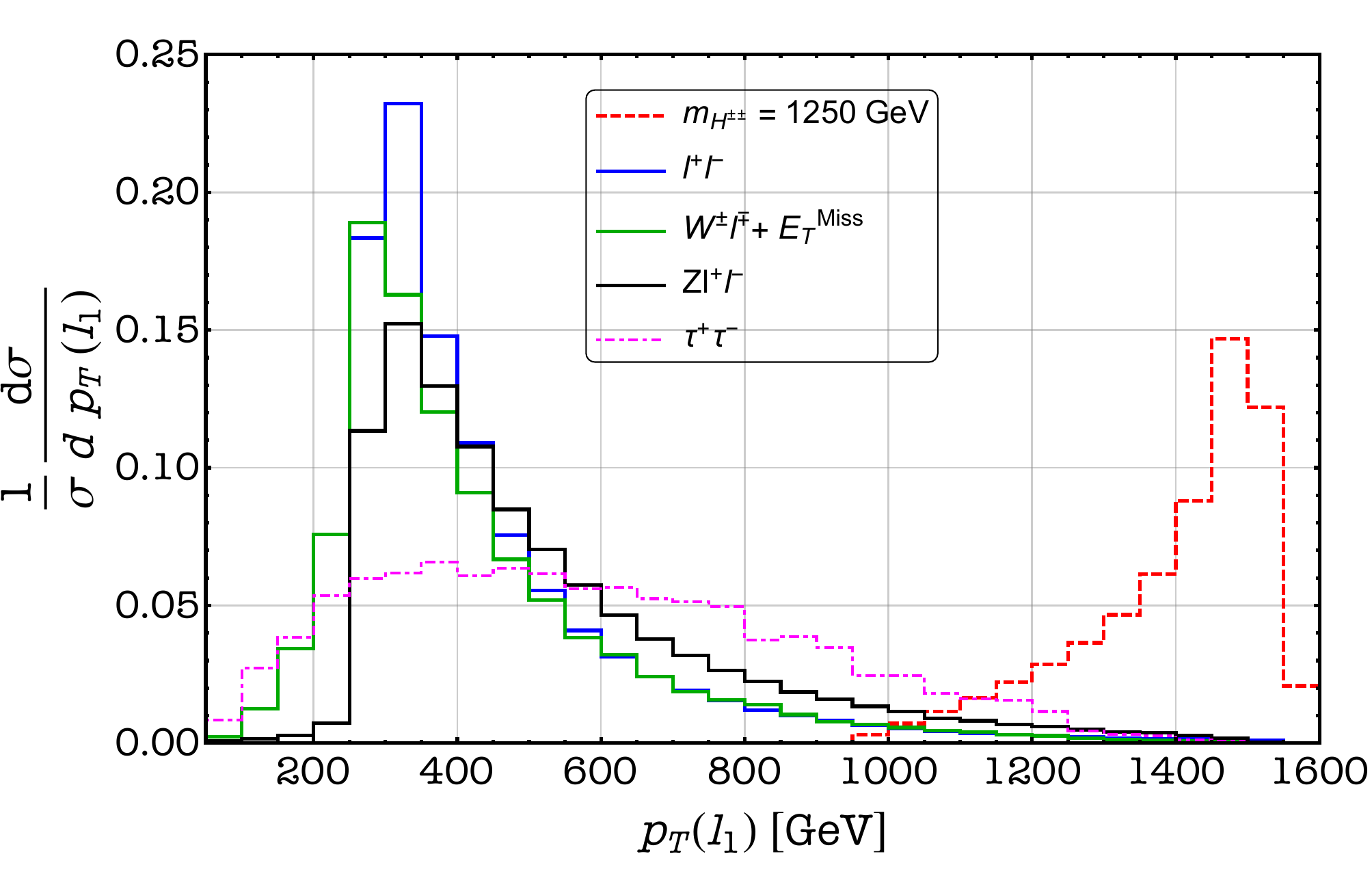}
\caption{Normalized distribution of kinetic variables $E^{\text{Miss}}_{T}$(left panel) and $p_T (\ell_1)$ (right panel) for both the signal and SM backgrounds for $\mu^+\mu^-$ final state.}
\label{Fig:mumuhist}
\end{figure}
A quick look at Table~\ref{Tab:SMBkg} suggests that most of the SM background subprocesses ($W \,\ell \, \nu$, $\tau^{+}\tau^{-}$ and $Z\ell^{+}\ell^{-}$) will all contain neutrinos in the final states from their decay products, which give rise to missing energy $\left(E^{\text{Miss}}_{T}\right)$ in the final state events. In contrast, both the signal and $\ell^{+}\ell^{-}$ background do not have any direct source of missing energy. 

In the \emph{left} panel of Figure \ref{Fig:mumuhist}, we show the $E^{\text{Miss}}_{T}$ distribution, whereas the \emph{right} panel illustrates the transverse momentum distribution of the leading muon, corresponding to both the signal and the relevant SM background processes. Note that, unlike the signal and $\ell^{+}\ell^{-}$ backgrounds, all the other backgrounds have a long tail in the missing energy distribution. Theoretically, for both the signal and $\ell^{+}\ell^{-}$ background, all the events must be concentrated in the bin with zero missing energy. However, due to the limitation in detector resolution efficiency, there is a finite mismeasurement probability that leads to a spread in the missing energy distribution for both these processes as well. To restrict the SM background that contains neutrinos and lead to large missing energy in the event topology, we impose $E^{\text{Miss}}_{T} < 100~\tt{GeV}$ to suppress their contributions. 
The effect of this cut becomes obvious as both the signal and $\ell^{+}\ell^{-}$ are marginally affected by this cut, while all the other SM backgrounds go down significantly. 
Table \ref{Tab:BPmumu1250}, illustrates the cut-flow chart for the signal-background analysis. 
Note that the $E^{\text{Miss}}_{T}$ cut hardly makes any improvement in the $\sigma_{S}/\sigma_B$ ratio (last column in the table) since the dominant background for this final state comes from the SM di-lepton production that contains the $\mu^{+}\mu^{-}$ final state. The transverse momentum of the leading muon $p_{T}\left(\mu_{1}\right)$ becomes instrumental in suppressing this large background, as shown in the \emph{right} panel of Figure \ref{Fig:mumuhist}. The leading muon's transverse momentum is seen to peak for large values (near $\tt{1450~GeV}$), whereas all the SM backgrounds peak for lower values and drop off around $\tt{1000~GeV}$. This is expected as the signal is mediated by a heavy propagator that leads to the final state leptons being produced more in the central rapidity region and therefore larger transverse momenta. On the other hand, for the dominant SM background arising from the $t$-channel exchange of photon and $Z$ boson, which are effectively massless propagators compared to the c.o.m. energy, the final state leptons have a stronger probability of being produced at small scattering angles and therefore smaller transverse momenta. Exploiting this feature in $p_{T}\left(\mu_{1}\right)$ distribution, we impose a cut of $p_{T}\left(\mu_{1}\right) > \tt{1350~GeV}$ that helps improve the signal significance $\sim \tt{2.5}$ times when compared to the $E^{\text{Miss}}_{T}$ cut\footnote{Instead of $p_{T}\left(\mu_{1}\right)$ one could also use a cut on $|\eta_{\mu_{1}}|$ to achieve similar significance as both these variables are highly correlated. We have compared the effects of both these cuts separately and find that the $p_{T}\left(\mu_{1}\right)$ cut performs better.}. 
\begin{table}[h!]
\centering
\begin{tabular}{|| c | c | c | c | c | c | c ||}
\hline
& $m_{H^{\pm\pm}} =  $ \tt{1250 GeV} & $\ell^{+}\ell^{-}$ & $W\ell\nu$ & $\tau^{+}\tau^{-}$ & $Z\ell^{+}\ell^{-}$ & $\frac{\sigma_{S}}{\sigma_{B}}$  \\
\hline
$\tt{N_{\mu} = 2}$ &  $\tt{3.59\times10^{2}}$ &$\tt{3.53\times10^{3}}$ & $\tt{5.82\times10^{2}}$ &\tt{2.60} & \tt{2.38} & \tt{0.087}  \\
\hline
\tt{$E^{\text{Miss}}_{T} <$ 100 GeV} &  $\tt{3.30\times10^{2}}$ & $\tt{3.52\times10^{3}}$ & $\tt{4.10\times10^{1}}$ & $\tt{0.61}$ & \tt{1.81} &  \tt{0.093} \\
\hline
\tt{$p_{T}\left(\mu_{1}\right) >$ 1350 GeV} &  $\tt{6.87\times10^{2}}$ & $\tt{2.00\times 10^{1}}$ & $\tt{7.64\times 10^{-2}}$ & \tt{0} & $\tt{0.01}$ & \tt{34.20} \\
\hline
\end{tabular}
\caption{Benchmark analysis for the $\mu\mu$ final state. All the cross-sections presented in the above table are in $fb$. In the last column, we use the signal-to-background ratio ($\frac{\sigma_{S}}{\sigma_{B}}$) to represent the efficiency of each cut.}
\label{Tab:BPmumu1250}
\end{table}

\subsection{\texorpdfstring{$e^{+}e^{-}$}{e+ e-} channel}
\begin{figure}[t!]
\centering
\includegraphics[scale=0.24]{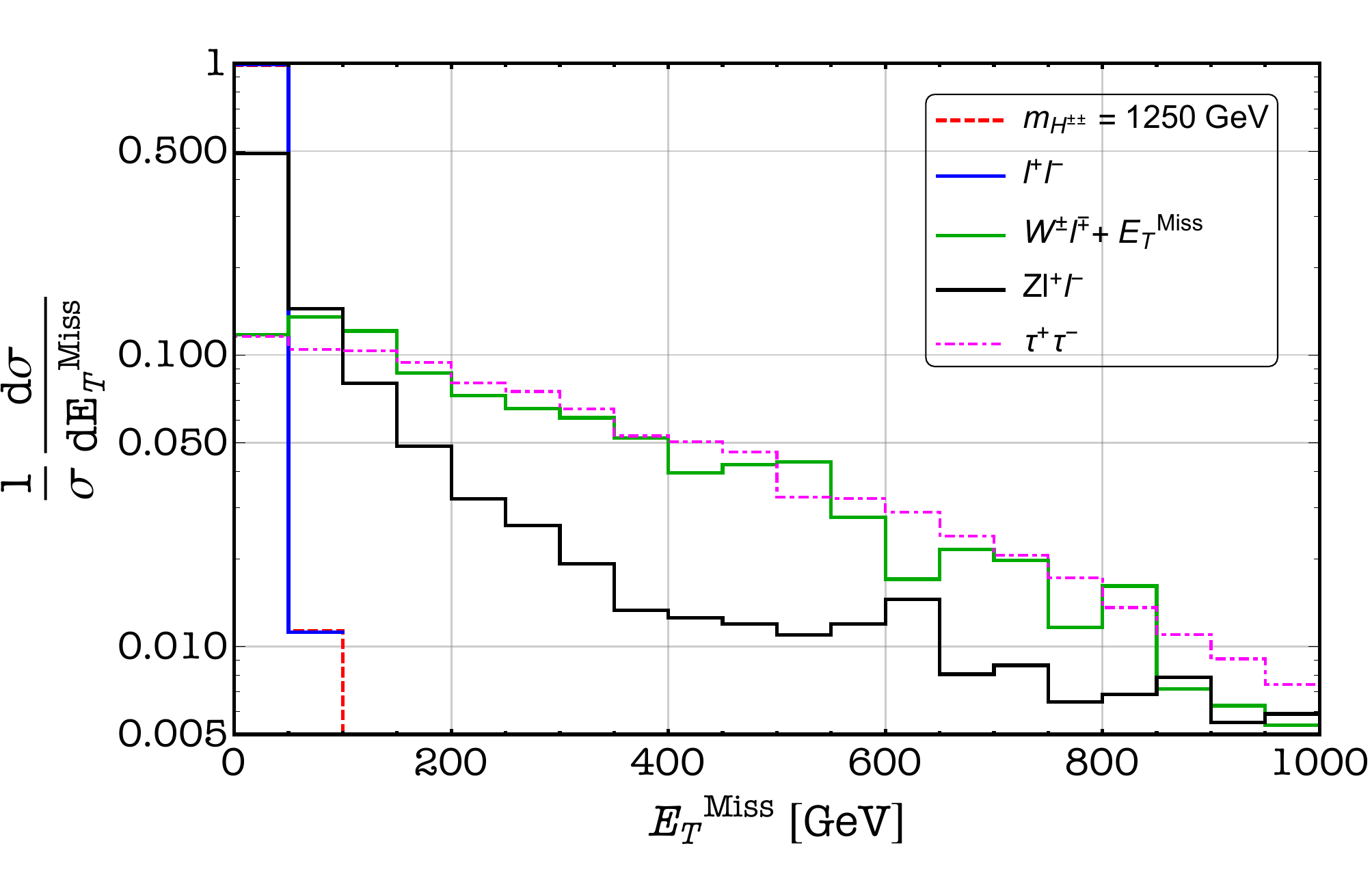}
\includegraphics[scale=0.24]{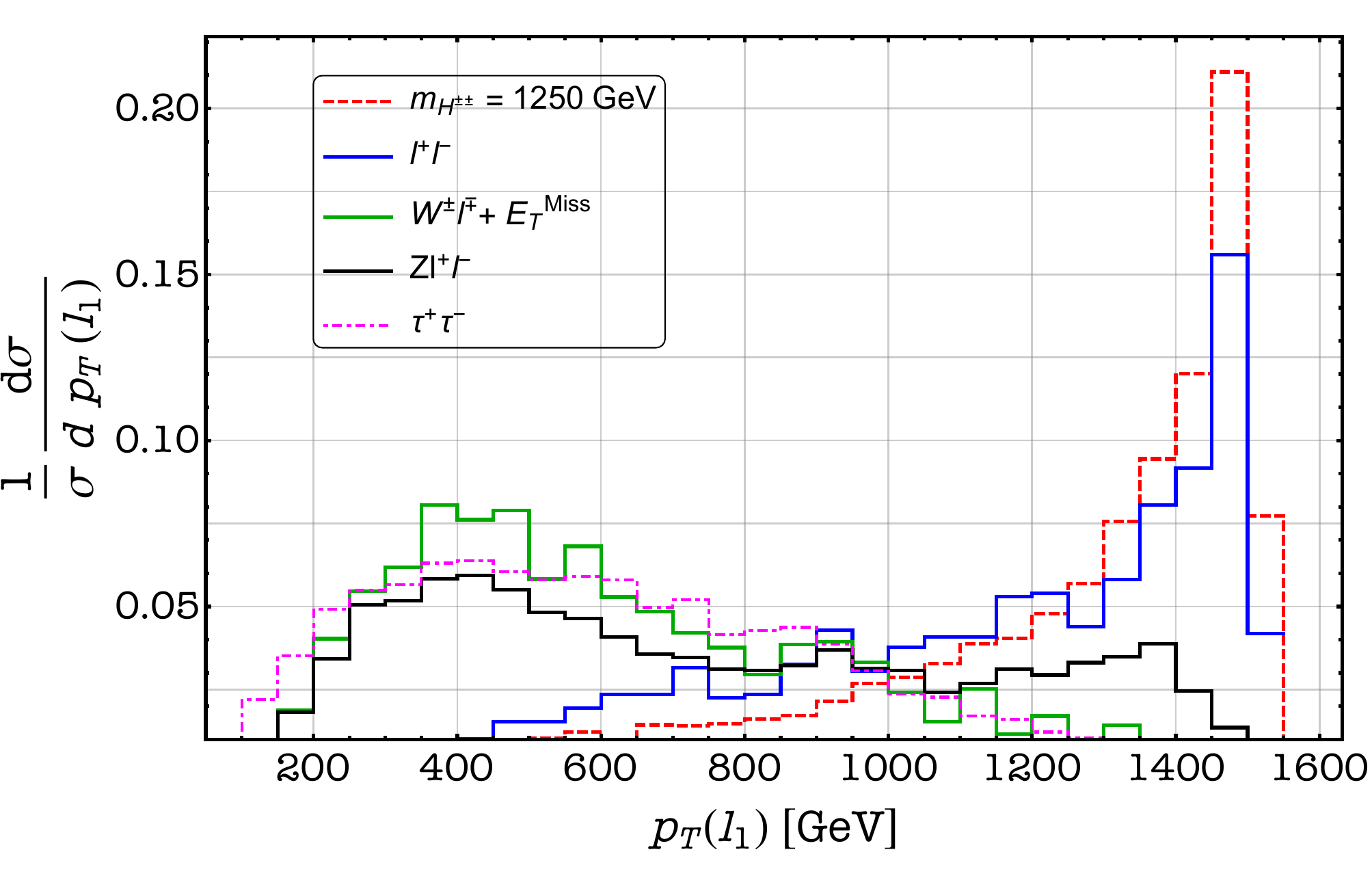}
\caption{Normalized distribution of kinetic variables $E^{\text{Miss}}_{T}$(left panel) and $p_T (\ell_1)$ (right panel) for both the signal and SM backgrounds for $e^+e^-$ final state.}
\label{Fig:eehist}
\end{figure}

For the $e^{+}e^{-}$ final state, we employ a similar strategy as described in the $\mu^{+}\mu^{-}$ case. However, after imposing $\tt{N_{e}} = 2$ selection criteria, the dilepton background no longer receives contributions from the $t$-channel, as previously discussed. This renders the dilepton ($\mu^+\mu^- \to e^+e^-$) background very small due to the inherent $s$-channel suppression. This can be seen in Table \ref{Tab:BPee1250} where, after demanding two electrons in the final state, the cross-section of the $\ell^{+}\ell^{-}$ background reduces significantly. As before, we choose a similar selection cut for the outgoing electrons, given by
\begin{equation}
\tt{p_{T}\left(e\right) > 50~GeV ,~~~ \left|\eta\left(e\right)\right| < 2.4} \,\,\, .
\label{Eq:eereconstruction}
\end{equation}
In Figure \ref{Fig:eehist}, we show the $E^{\text{Miss}}_{T}$ and $p_{T}\left(e_{1}\right)$ distributions for the signal and SM background processes. Similar to the case of the muon final states, all SM contributions exhibit significant events at large $E^{\text{Miss}}_{T}$ except for the di-lepton background. The signal and di-lepton background events are again concentrated within the $\tt{100~GeV}$ bin. However, the missing transverse energy cut proves to be more effective for this final state, as the remaining dilepton background, being largely unaffected by this cut, is already minimal. Therefore, the overall cross-section for the SM background is significantly reduced by this cut.
In Table \ref{Tab:BPee1250}, we show the cut flow chart for the $e^{+}e^{-}$ final state. We first impose an upper bound on $E^{\text{Miss}}_{T} < \tt{50~GeV}$, which now enables us to achieve an improved $\sigma_{S}/\sigma_B$ ratio. 
We observe that the $p_{T}\left(e_{1}\right)$ distribution for the di-lepton background is unlike the muon case, as it no longer has the substantial contributions from the $t$-channel process. Since the dominant production mechanism for the $e^+e^-$ final state is an $s$-channel process, the outgoing fermions tend to be produced back-to-back in the central region, leading to a characteristic peak around half the c.o.m. energy, as illustrated in the figure. Fortunately, the dilepton background is small in this case, and therefore, the application of the cut $p_{T}\left(e_{1}\right) > \tt{1300}~\tt{GeV}$ significantly suppresses the remaining background contributions, effectively doubling the signal-to-background ratio. 
\begin{table}[ht!]
\centering
\begin{tabular}{|| c | c | c | c | c | c | c ||}
\hline
& $m_{H^{\pm\pm}} =  $\tt{1250 GeV} & $\ell^{+}\ell^{-}$ & $W\ell\nu$ & $\tau^{+}\tau^{-}$ & $Z\ell^{+}\ell^{-}$ & $\frac{\sigma_{S}}{\sigma_{B}}$  \\
\hline
$\tt{N_{e} = 2}$ & $\tt{1.27\times10^{3}}$ &$\tt{8.67}$ & $\tt{21.32}$ &\tt{2.45} & $\tt{5.82\times10^{-2}}$ & \tt{39.08}  \\
\hline
\tt{$E^{\text{Miss}}_{T} <$ 50 GeV} &  $\tt{1.26\times10^{3}}$ & $\tt{8.58}$ & $\tt{2.50}$ & $\tt{0.28}$ & $\tt{2.86}\times10^{-2}$ & \tt{110.64} \\
\hline
\tt{$p_{T}\left(e_{1}\right) >$ 1300 GeV} &  $\tt{7.31\times10^{2}}$ & $\tt{3.62}$ & $\tt{0.04}$ & $\tt{5.0\times10^{-4}}$ & $\tt{3.89\times10^{-4}}$ &  \tt{199.68} \\
\hline
\end{tabular}
\caption{Benchmark analysis for the $e^+e^-$ final state. All the cross-sections presented in the above table are in $fb$. In the last column, we use the signal-to-background ratio ($\frac{\sigma_{S}}{\sigma_{B}}$) to represent the efficiency of each cut.}
\label{Tab:BPee1250}
\end{table}

\subsection{\texorpdfstring{$\tau^{+}\tau^{-}$}{tau+ tau-} channel}
For the $\tau^{+}\tau^{-}$ final state, we adopt a different strategy owing to the challenges associated with the $\tau$ reconstruction. Here, for our analysis, we demand that both the $\tau$ in the final state must decay via leptonic mode\footnote{One can also choose hadronic decays of the $\tau$. However, in that scenario, one must also account for an additional set of jet-associated SM background.}. Among the three possible pure leptonic final states (\emph{i.e.} $\mu^{+} \mu^{-} \to e^{+} e^{-} + E^{\text{Miss}}_{T}$, $\mu^{+} \mu^{-} \to \mu^{+} \mu^{-} + E^{\text{Miss}}_{T}$, $\mu^{+} \mu^{-} \to \mu^{\pm}e^{\mp} + E^{\text{Miss}}_{T}$), we demand that the final state consists of $\tt{N_{e} = 1~\&~N_{\mu} = 1}$, maintaining the kinematic requirements of Eq.(\ref{Eq:muconstract}) and Eq.(\ref{Eq:eereconstruction}). 
This allows us to avoid the dominant SM background associated with the $t$-channel dilepton production.

\begin{figure}[t!]
\centering
\includegraphics[scale=0.2]{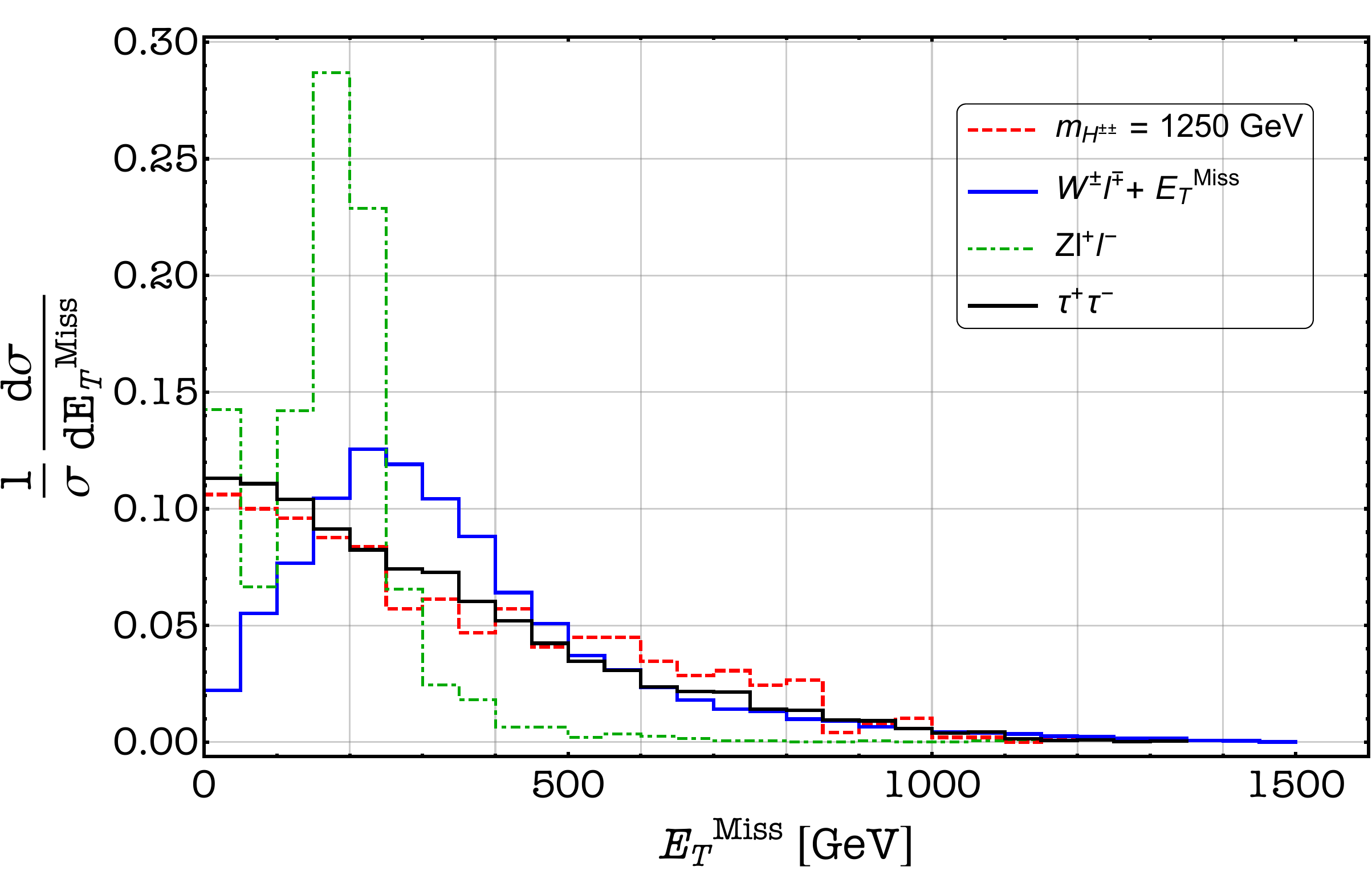}
\includegraphics[scale=0.2]{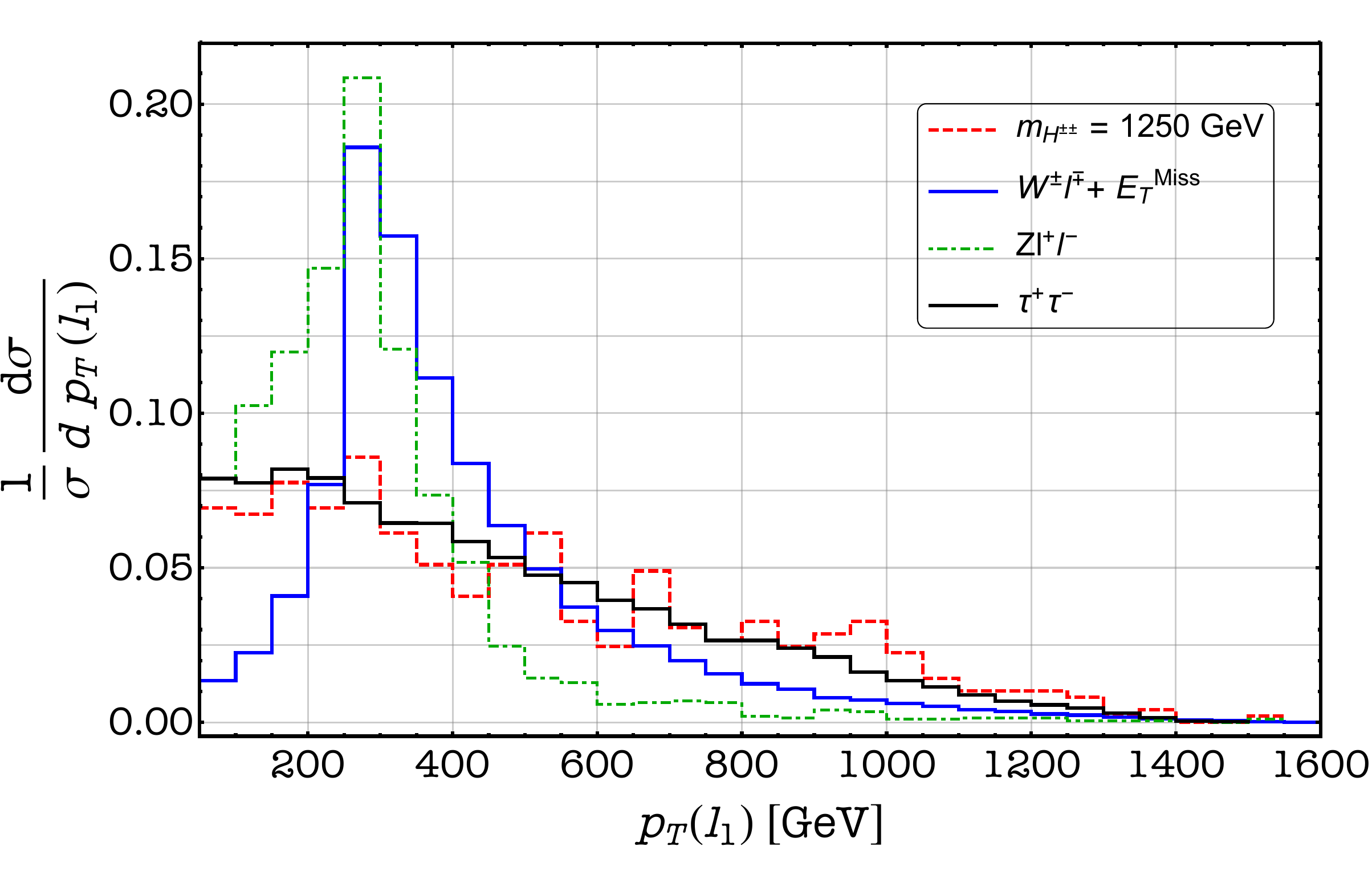}
\caption{Normalized distribution of kinetic variables $E^{\text{Miss}}_{T}$(left panel) and $p_T (\ell_1)$ (right panel) for both the signal and SM backgrounds for $\tau^+\tau^-$ final state.}
\label{Fig:tatahistpTeT}
\end{figure}
In Figure \ref{Fig:tatahistpTeT} we display the distributions corresponding to $E^{\text{Miss}}_{T}$ and $p_{T}(\ell_{1})$\footnote{We wish to point out that in this case leading lepton can be either electron or muon.} for both the signal and the backgrounds. From the shape of the distributions, one can ascertain that there is no specific discriminating feature that can help in devising a cut to improve the signal significance. Interestingly, we find that the signal distributions in $\tau$ completely lose the characteristic peak in the transverse momentum that was observed for both the $\mu\mu$ and $ee$ channels earlier, following a similar production mode. Hence, we consider alternative variables. 
%
\begin{figure}[h!]
\centering
\includegraphics[scale=0.25]{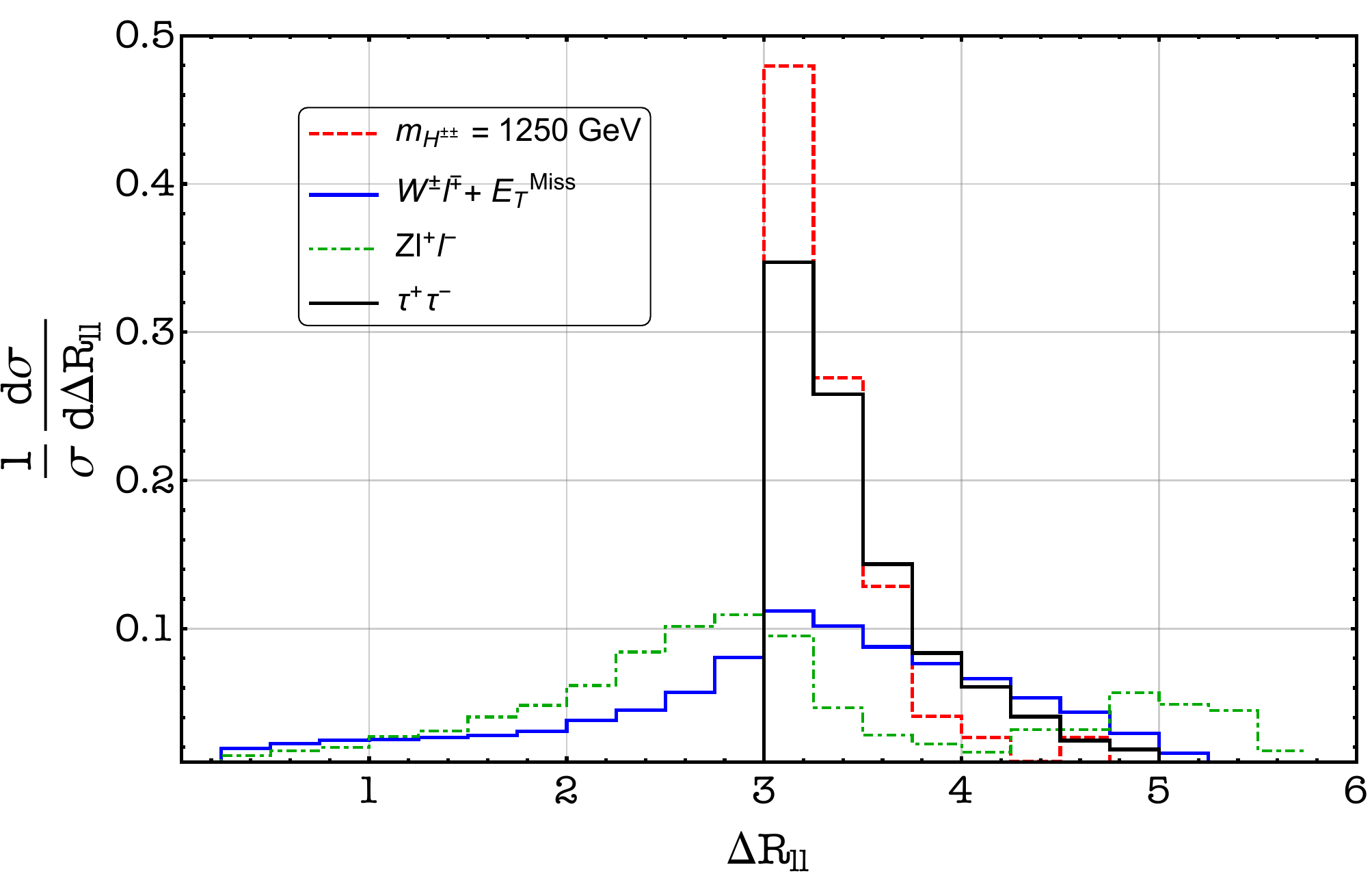}
\includegraphics[scale=0.205]{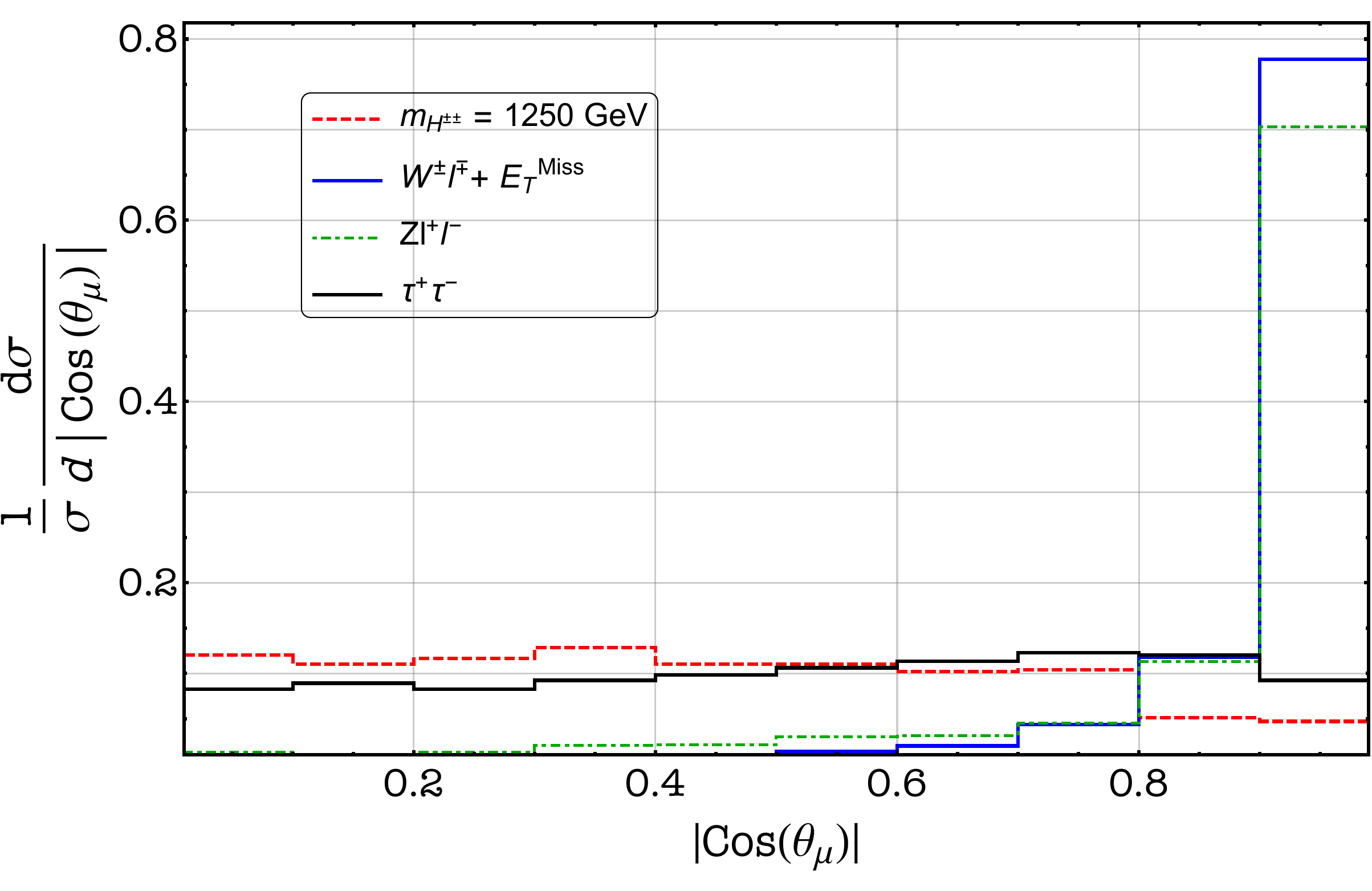}
\caption{Normalized distribution of kinetic variables $\Delta R_{\ell\ell}$(left panel) and $\cos\theta_\mu$ (right panel) for both the signal and SM backgrounds for $\tau^+\tau^-$ final state.}
\label{Fig:tatahist}
\end{figure}
As absolute energy variables do not seem to help our cause, we consider angular variables to check whether we find any distinguishing features between the signal and SM backgrounds. 
In Figure~\ref{Fig:tatahist} we show the distributions in $\Delta R_{\ell\ell}$, which measures the angular separation between the two outgoing leptons, and $\tt{\left|\cos\theta_{\mu}\right|}$ which represents the angle of the $\mu$-tagged lepton with the beam axis, for the signal and SM background events.
We observe that the variable $\Delta R_{\ell\ell}$ exhibits a pronounced peak for both the signal and the $\tau^+\tau^-$ background, whereas other background processes show a more uniform distribution across the full range. Additionally, the event shape of the absolute value of the complementary variable $\tt{\left|\cos\theta_{\mu}\right|}$ reveals that the $W\ell\nu$ and $Z\ell^{+}\ell^{-}$ backgrounds predominantly peak in the $\tt{0.9}$ to $\tt{1.0}$ bin as opposed to the signal and the $\tau^{+}\tau^{-}$ background. This motivates the application of the selection cut $\tt{|\cos\theta_{\mu}| < 0.8}$ to suppress these backgrounds, further improving the signal-to-background ratio. 
In Table \ref{Tab:BPtata1250}, we present the details of the cut and count analysis. 
\begin{table}[h!]
\centering
\begin{tabular}{|| c | c | c | c | c | c ||}
\hline
& $m_{H^{\pm\pm}} =  $\tt{1250 GeV}  & $W\ell\nu$ & $\tau^{+}\tau^{-}$ & $Z\ell^{+}\ell^{-}$ & $\frac{\sigma_{S}}{\sigma_{B}}$  \\
\hline
$\tt{N_{e} = 1~\&~N_{\mu} = 1}$  &  $\tt{8.17\times10^{1}}$ & $\tt{5.35\times10^{2}}$ &\tt{5.37} & $\tt{2.31\times10^{-2}}$ & \tt{0.15}  \\
\hline
\tt{$\left|\cos\theta_{\mu}\right| <$ 0.8} &  $\tt{7.34\times10^{1}}$ & $\tt{5.61\times10^{1}}$ & $\tt{4.22}$ & $\tt{4.26\times10^{-3}}$ &  \tt{1.22} \\
\hline
\end{tabular}
\caption{Benchmark analysis for the $\tau\tau$ final state. All the cross-sections presented in the above table are in $fb$.}
\label{Tab:BPtata1250}
\end{table}

\section{Results}\label{sec:summ}
\begin{figure}[t!]
\centering
\includegraphics[scale=0.4]{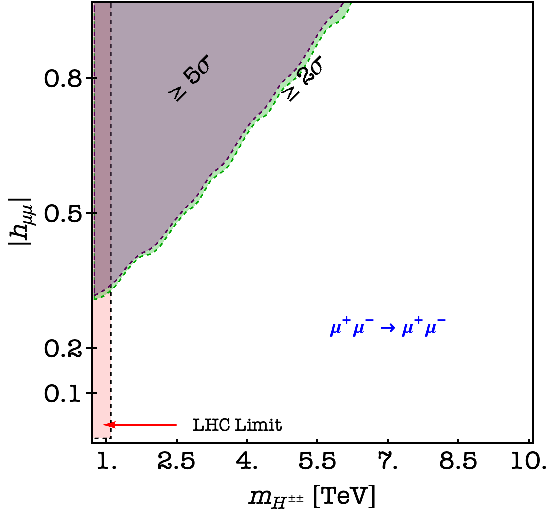}
\caption{Illustrating the sensitivity reach for $|h_{\mu\mu}|$ as a function of the doubly charged scalar mass following the cut and count analysis described in section \ref{sec:analysis}, for an integrated luminosity $\mathcal{L}=\tt{1~ab^{-1}}$ and $\tt{\sqrt{s}= 3~TeV}$.}
\label{Fig:bounds1}
\end{figure}

\begin{figure}[h]
\centering
\includegraphics[scale=0.4]{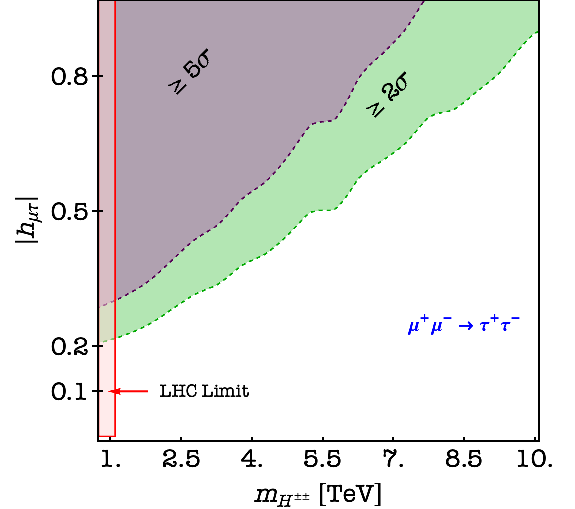}
\caption{Illustrating the sensitivity reach for $|h_{\mu\tau}|$ as a function of the doubly charged scalar mass following the cut and count analysis described in section \ref{sec:analysis}, for an integrated luminosity $\mathcal{L}=\tt{1~ab^{-1}}$ and $\tt{\sqrt{s}= 3~TeV}$.}
\label{Fig:bounds3}
\end{figure}

We now demonstrate 
how the muon collider with $\sqrt{s} = 3~\tt{TeV}$, using the three final state topologies $\mu^{+}\mu^{-}$, $e^{+}e^{-}$ and $\tau^{+}\tau^{-}$, improves our sensitivity reach for the mass and leptonic couplings of the doubly-charged scalar. The obtained results, following the methodology discussed in the previous section, enable us to determine the maximum sensitivity reach for each Yukawa coupling involving the doubly charged scalar. The sensitivity plots for the three cases analyzed in the previous section in the coupling-mass plane of the parameter space are shown in Figures~\ref{Fig:bounds1}, \ref{Fig:bounds2}, and \ref{Fig:bounds3}. For the sensitivity plots, we have chosen an integrated luminosity $\mathcal{L}=\tt{1~ab^{-1}}$. 
We discuss below our findings in each of the sensitivity plots.

\begin{itemize}
\item \textbf{$\tt{\mu^{+}\mu^{-}} \to \mu^{+}\mu^{-}$ :}

We have previously outlined the event selection criteria to enhance signal significance for the 
$\mu^+\mu^-$ final state. Since the signal cross-section was initially generated assuming $h_{\mu\mu} = 1$, it can be straightforwardly rescaled for any given value of $|h_{\mu\mu}|$ corresponding to different scalar mass choices. The variation of the signal cross-section with respect to the scalar mass is already shown in Figure~\ref{Fig:sigxsec}.
In Figure~\ref{Fig:bounds1}, we illustrate the parameter space in the ${m_{H^{\pm\pm}}, |h_{\mu\mu}|}$ plane that can be probed using the cut-flow strategy detailed in Table~\ref{Tab:BPmumu1250}. The \emph{dark green} and \emph{gray} shaded regions represent the areas where a statistical significance exceeding $\tt{2\sigma}$ and $\tt{5\sigma}$, respectively, can be achieved. The \emph{light-red} vertical bar represents the direct search limit from $\tt{LHC}$ ($m_{H^{\pm\pm}} \lesssim \tt{1100~GeV}$) extracted from the $\tt{ATLAS}$ analysis\cite{ATLAS:2022pbd}
It is evident from the figure that our analysis substantially extends the discovery reach for the doubly charged scalar mass $m_{H^{\pm\pm}}$, from the current exclusion limit on $m_{H^{\pm\pm}}$, 
\begin{figure}[b!]
\centering
\includegraphics[scale=0.4]{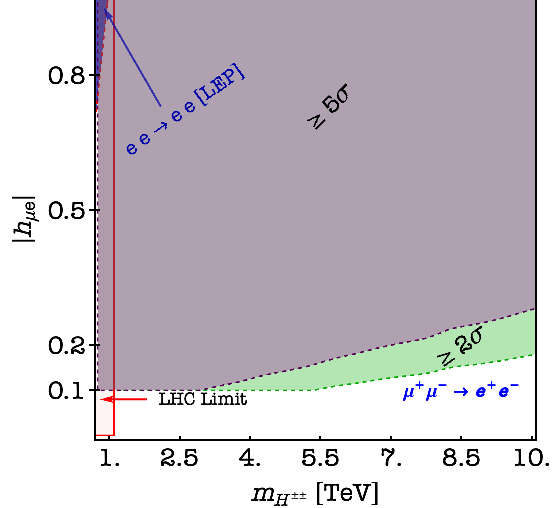}
\caption{Illustrating the sensitivity reach for $|h_{\mu e}|$ as a function of the doubly charged scalar mass following the cut and count analysis described in section \ref{sec:analysis}, for an integrated luminosity $\mathcal{L}=\tt{1~ab^{-1}}$ and $\tt{\sqrt{s}= 3~TeV}$.}
\label{Fig:bounds2}
\end{figure}
for Yukawa couplings in the range $\left|h_{\mu\mu}\right| \approx \tt{0.15}$ to $\left|h_{\mu\mu}\right| \approx \tt{0.65}$. In fact, no direct limits on the $h_{\mu\mu}$ coupling exist in the literature, and the muon collider will provide an exclusive opportunity to probe this coupling to very small values for an extended range of doubly charged scalar mass beyond the reach of LHC and lepton-flavor-violation experiments.  
\item \textbf{$\tt{\mu^{+}\mu^{-}} \to e^{+}e^{-}$ :}

In this channel, the contribution to the signal comes through the $t$-channel exchange of the doubly charged scalar, and the vertex violates lepton flavor. This process, therefore, corresponds to the flavor-violating vertex $h_{\mu e}$. 
In Figure \ref{Fig:bounds2} we illustrate the region in the parameter plane $\{m_{H^{\pm\pm}}, \left|h_{\mu e}\right|\}$ which can be probed using the cut-flow chart discussed in 
Table.~(\ref{Tab:BPee1250}). 
The \emph{dark green} and \emph{gray} shaded regions in the plot indicate areas of the parameter space that can be probed with statistical significances exceeding $2\sigma$ and $5\sigma$, respectively. Additionally, the \emph{dark blue} shaded area in the upper-left corner corresponds to the $e^+e^- \to e^+e^-$ constraint from LEP, as summarized in Table~\ref{Tab:flavour}. The \emph{light red} shaded region represents the existing $\tt{LHC}$ limit, as discussed previously. The reach plot suggest that, one can probe the $m_{H^{\pm\pm}}$  in the $\tt{1.1~TeV}$ to $\tt{10~TeV}$ range for the corresponding $h_{\mu e}$ $\approx \tt{0.1}$ to $h_{\mu e}$ $\approx \tt{0.41}$. When compared with current experimental bounds, the reach of our analysis represents a substantial improvement over both collider and flavor experiments.   
\item  \textbf{$\tt{\mu^{+}\mu^{-}} \to \tau^{+}\tau^{-}$ :}

Finally, we comment on the significance of the 
$\tau^+\tau^-$ final state in probing the flavor violating $h_{\mu\tau}$ coupling. In Figure \ref{Fig:bounds3} we show the region of the parameter plane $\{m_{H^{\pm\pm}}, \left|h_{\mu \tau}\right|\}$ which can be explored following the cut flow chart described in Table \ref{Tab:BPtata1250}. For this signal topology, the most stringent available bound is from 
$\tt{LHC}$, which is illustrated as \emph{light-red} vertical bar. In this case as well, we show the $2\sigma$ and (\emph{dark-green}) $5\sigma$ (\emph{gray}) discovery contours, respectively. Compared to the previous two 
scenarios ($\mu\mu$ and $ee$), both the exclusion and discovery mass reaches are noticeably lower and reach values of $m_{H^{\pm\pm}} \lesssim \tt{8~TeV}$ at $2\sigma$ and $m_{H^{\pm\pm}} \lesssim \tt{6~TeV}$ at $5\sigma$ for $h_{\mu \tau} \approx \mathcal{O}\left(1\right)$. This is because of the challenges in $\tau$ identification, which significantly reduces the signal events. Notwithstanding this fact, the muon collider will still be able to probe the $h_{\mu\tau}$ coupling and the corresponding mass of the doubly charged scalar beyond any other foreseeable experiment.    
\end{itemize}

\section{Resolving a Possible Inverse Problem} 
\label{sec:inverse}
Before summarizing our primary findings, we would like to highlight an interesting possibility regarding the leptonic channels under consideration. As charge conservation is implicit, the analyzed final states can appear in several different incarnations of a BSM scenario. Of the many likely options of different spin particles contributing to the same process, the most notable option to highlight here would be the exchange of the neutral scalar that could mediate the processes in the same way as the doubly charged scalar. As the spin of the exchanged mediator has the knack of showing up through characteristic angular dependence in the final state particles, the challenge is more pronounced if we consider the same spin particle as an alternative. The central question is whether a muon collider possesses the capability to distinguish between different underlying scenarios, a challenge that lies at the heart of the \emph{inverse problem} in particle physics.

\begin{figure}[ht!]
\centering
\includegraphics[height=3.5cm,width=6.5cm]{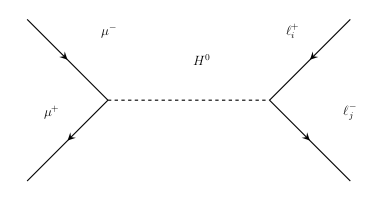}
\hspace{0.35cm}
\includegraphics[height=3.5cm,width=6.5cm]{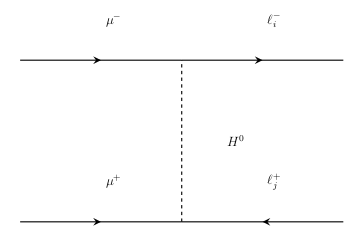}
\caption{Illustrating the $s$- and $t$-channel Feynman diagrams corresponding to the $H^{0}$ mediated signal process $\mu^{+} \mu^{-} \to \ell^{+}_{i} \ell^{-}_{j}$, where the index $i, j = e, \mu, \tau$.}
\label{Fig:Feyn_Neutral}
\end{figure}

To elaborate this, let us consider an alternative BSM scenario where a $\tt{CP}$-even heavy neutral scalar $H^{0}$ can give rise to identical signal topology $\mu^{+}\mu^{-} \to \ell^+_{i}\ell^-_{j}$ instead of a doubly charged scalar which we have considered above. The interaction Lagrangian for this substitute new physics model takes a very similar form to what we consider for the doubly charged scalar in Eq.~(\ref{Eqn:theory}), given as
\begin{equation}
\mathcal{L}_{H^{0}} \supset \sum_{i,j = e,\mu, \tau}\tilde{h}_{ij}\overline{\ell}_{i}H^{0}\ell_{j}.
\label{Eq:ModelH0}
\end{equation}   
Here, all the fields are expressed in the mass basis, and the coefficient $\tilde{h}_{ij}$ represents the coupling strength of the $\tt{CP}$-even neutral scalar $H^{0}$ with the SM charged leptons. Based upon the model, one can write down the explicit form of the leptonic coupling $\tilde{h}_{ij}$, which is a function of different parameters of the underlying model. For example, in case of Type-I 2HDM \cite{Branco:2011iw}, the coupling $\tilde{h}_{ii} = \frac{m_{\ell_{i}}}{v}\cot\beta$ (the parameter $\beta$ signifies the ratio between two \emph{vev}'s of the model) and the coupling $\tilde{h}_{ij}~\text{with}~i \neq j$ is zero as the model does not allow $\tt{FCNC}$. However, in models such as the Type-III 2HDM and the MSSM, neutral scalars can contribute to FCNC processes at tree level. In the present work, we consider a framework in which neutral scalars do not induce tree-level FCNCs. The interaction term of Eq.(\ref{Eq:ModelH0}) can give rise to $\mu^{+}\mu^{-} \to \ell^+_{i}\ell^-_{j}$ processes and the corresponding Feynman diagrams are shown in Figure~\ref{Fig:Feyn_Neutral}. From these diagrams, it is obvious that the final state arising from the neutral scalar-mediated process closely resembles that of the $H^{\pm\pm}$-mediated process. (see Figure~\ref{Fig:feynman} for comparison).   
In this situation, one can ask the following question - 
\begin{itemize}
\item[] Suppose after analyzing the $\mu^{+} \mu^{-} \to \mu^{+} \mu^{-}$ final state in the future muon collider, one observes significant excesses of events \emph{w.r.t} the $\tt{SM}$ background. Given the circumstances, can one infer the nature of the underlying new physics (in this case, whether the excess is from a neutral or doubly charged scalar) responsible for the excess? 
\end{itemize}
This is a crucial argument in what is termed as the ``\emph{Inverse problem}'' and efforts exist \cite{Bhattacherjee:2008ik} to address such questions in the context of a future $e^{+}e^{-}$ collider. Before introducing a discriminating variable aimed at resolving this issue, we would like to underscore the differences between Figure~\ref{Fig:feynman} and Figure~\ref{Fig:Feyn_Neutral}, respectively. As opposed to the doubly charged scalar-mediated process involving $H^{\pm\pm}$, the neutral scalar scenario includes an $s$-channel mediated contribution. The presence of this diagram can substantially contribute to the corresponding cross-section. However, if one considers a neutral scalar mass $m_{H^{0}}$ that is sufficiently far away from the $\tt{c.o.m}$ energy of the collider, then one can safely ignore its contributions. Under this premise, we are now in a position to compare two $t-channel$ processes (see Figure~\ref{Fig:feynman} and \emph{left} panel of Figure~\ref{Fig:Feyn_Neutral}) which emerge from two distinct model hypothesis $\mathcal{L}_{H^{\pm\pm}}$ (see Eq.(\ref{Eqn:theory})) and $\mathcal{L}_{H^{0}}$ (see Eq.(\ref{Eq:ModelH0})), respectively. The striking distinction between these two diagrams is that, for the $H^{\pm\pm}$ mediated process, there is a charge flip in the final state muon leg, which does not occur in the case of the heavy neutral scalar $H^{0}$. This feature arises due to the presence of the charge conjugation operator in the $\mathcal{L}_{H^{\pm\pm}}$ Lagrangian. This aspect acts as the \emph{de facto} rationale to build up the observable, which in turn will answer the problem at hand. 

To address the aforementioned question, we simulate the $\mu^{+}\mu^{-} \to \mu^{+}\mu^{-}$ final state for these two different \emph{new physics} scenarios separately using $\tt{MadGraph5@NLO}$. We set the corresponding masses and couplings as $m_{H^{\pm\pm}} = m_{H^{0}} = 1250~\tt{GeV}$ and $h_{\mu\mu} = \tilde{h}_{\mu\mu} = 1$ \footnote{Here we demand both these couplings $h_{\mu\mu},~\tilde{h}_{\mu\mu}$ are 1. To achieve that coupling strength, one needs to ensure that the neutral and the doubly charged scalar do not belong to the same scalar multiplet. This restriction arises from the hypercharge assignment of the scalar multiplet}. 
\begin{figure}[ht!]
\centering
\includegraphics[scale=0.31]{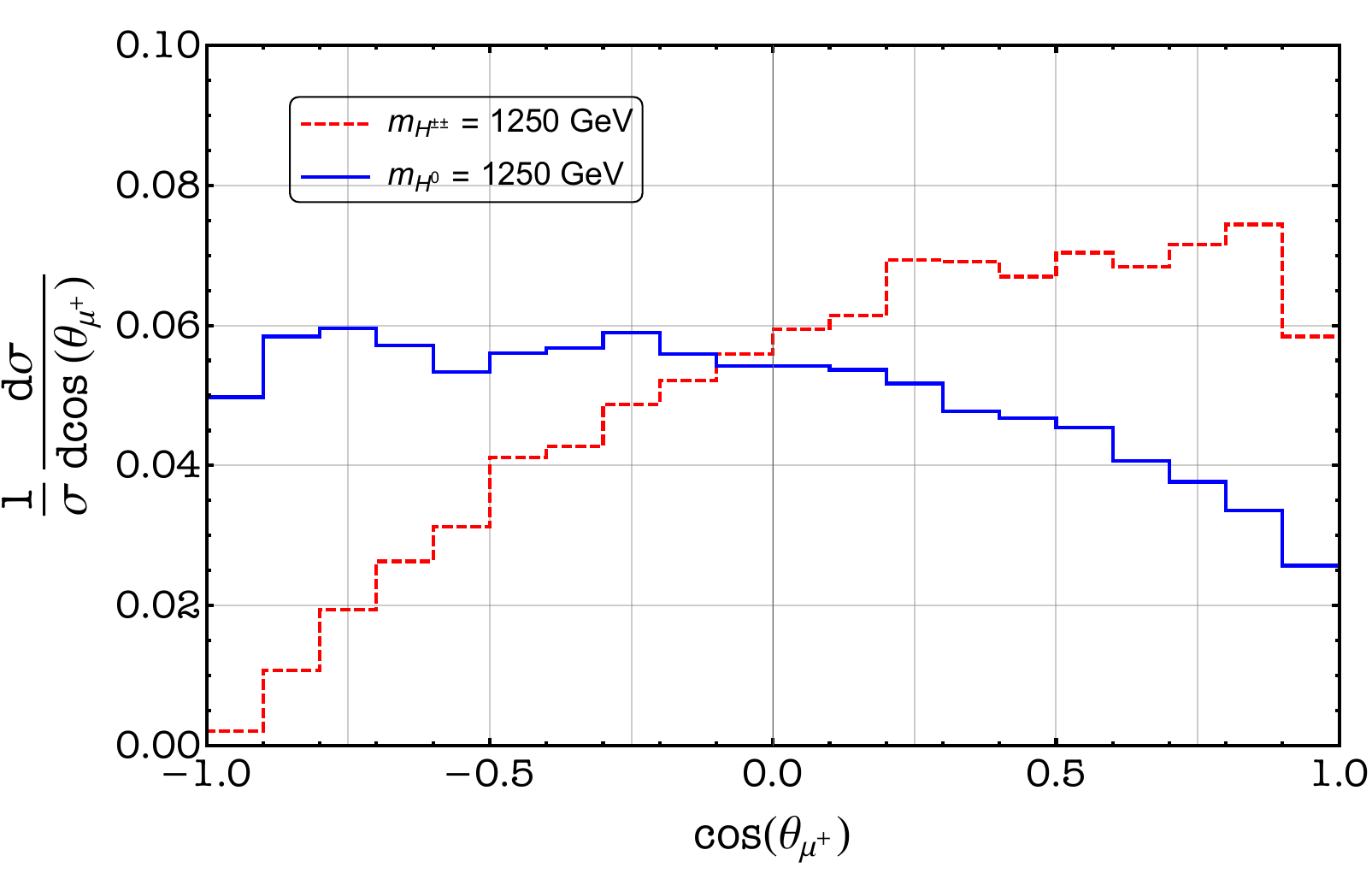}
\includegraphics[scale=0.31]{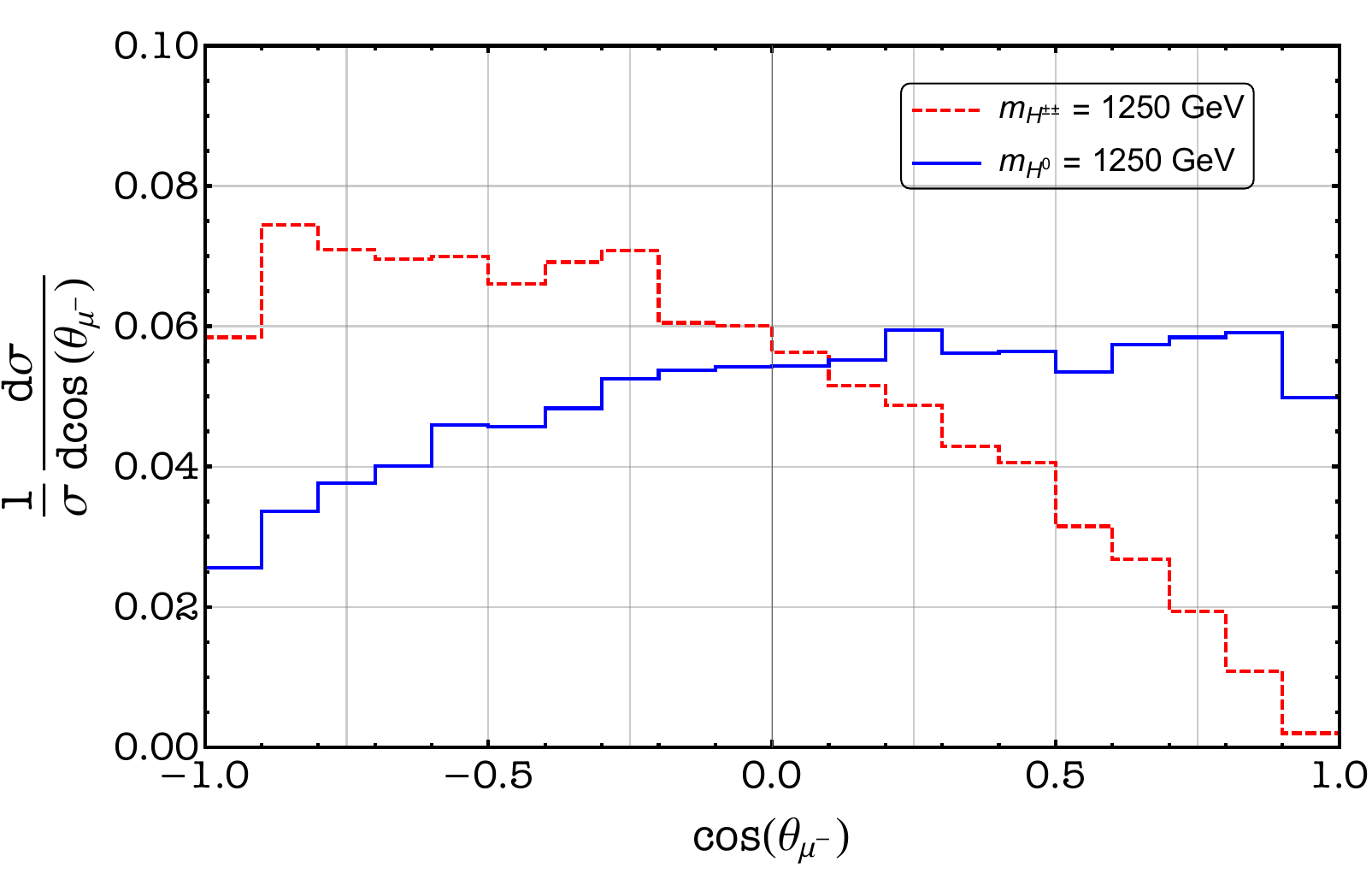}
\caption{Normalized distribution of angular variables $\cos\left(\theta_{\mu^{+}}\right)$ (\emph{left} panel) and $\cos\left(\theta_{\mu^{-}}\right)$ (\emph{right} panel) for two different hypothesis $\mathcal{L}_{H^{\pm\pm}}$ and $\mathcal{L}_{H^{0}}$.}
\label{Fig:H0vsHpp}
\end{figure}

\begin{figure}[ht!]
\centering
\includegraphics[width=0.48\textwidth]{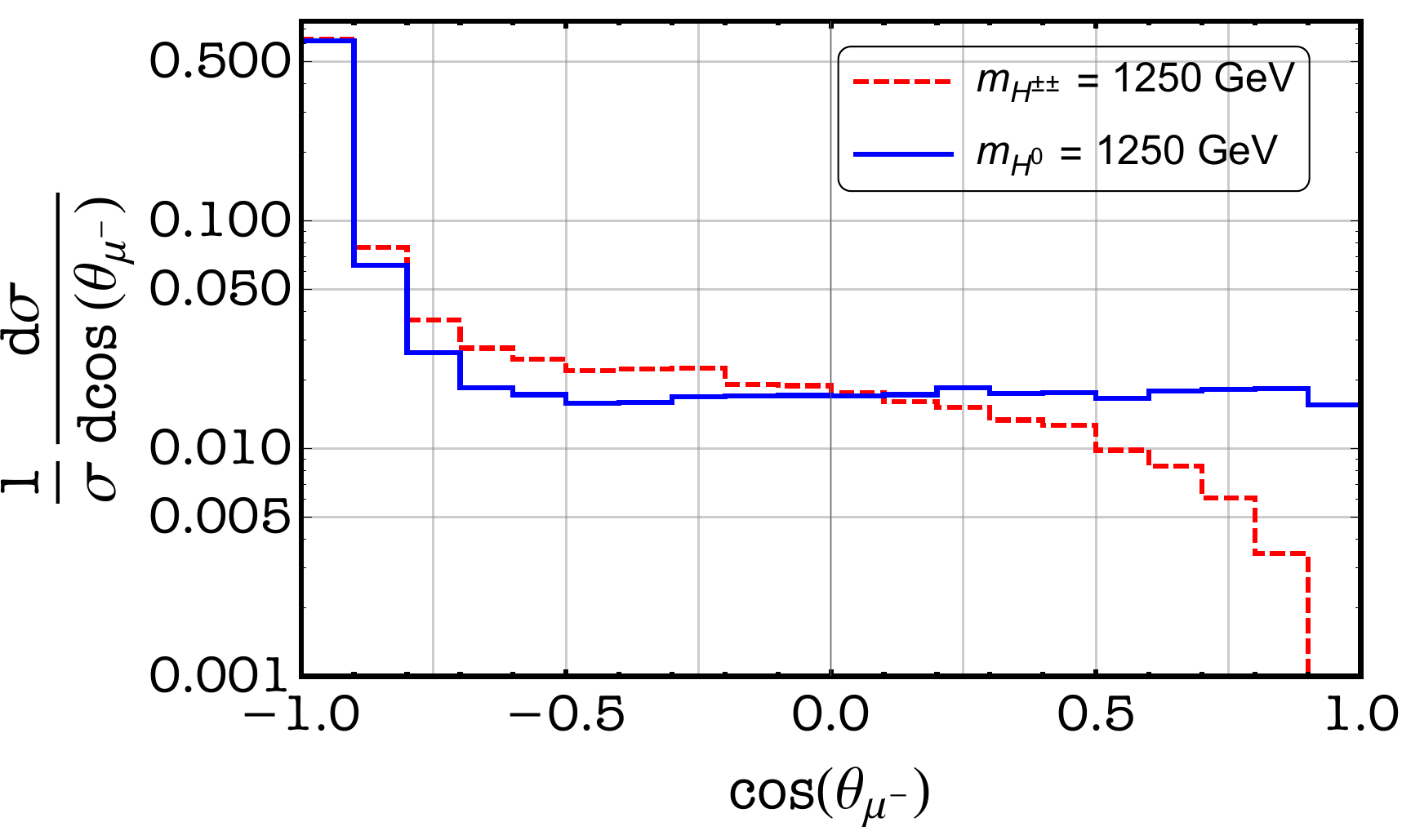}
\hfill
\includegraphics[width=0.48\textwidth]{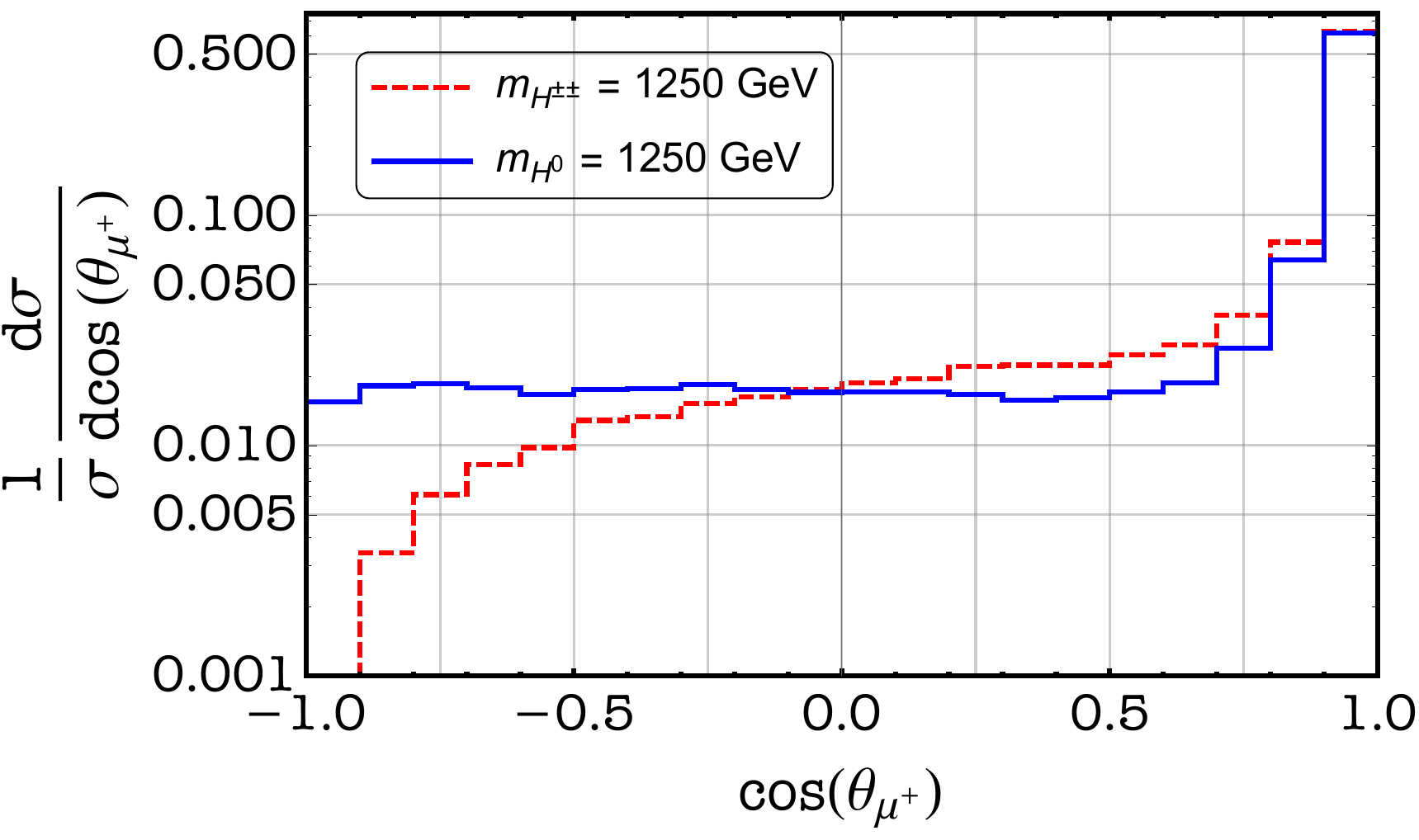}
\caption{Normalized distribution of angular variables $\cos\left(\theta_{\mu^{+}}\right)$ (\emph{left} panel) and $\cos\left(\theta_{\mu^{-}}\right)$ (\emph{right} panel) for two different hypotheses $\mathcal{L}_{H^{\pm\pm}}$ and $\mathcal{L}_{H^{0}}$. The distributions include both signal and background contributions.}
\label{Fig:H0vsHppWbkg}
\end{figure}

To analyze both these signals, we demand, emphasizing the charge identification of the muon, that $\tt{N_{\mu^{+}}} = \tt{N_{\mu^{-}}} = \tt{1}$, using the criteria given in Eq.(\ref{Eq:muconstract}). Our goal here is to construct a detectable observable that can enable us to distinguish between these two different new physics signals. To do so, we take into account the charge flipping of the final state muon leg, which happens for the doubly charged scalar scenario. As a result, one can think of angular variables $\cos\left(\theta_{\mu^{+}}\right)$ and $\cos\left(\theta_{\mu^{-}}\right)$. The corresponding distributions for these variables are presented in Figure~\ref{Fig:H0vsHpp}. In case of $\cos\left(\theta_{\mu^{+}}\right)$ distribution, one can notice that for the doubly charged case, the events are monotonously increasing over the range $\left[-1, 1\right]$. In contrast, for the neutral scalar case, the events are monotonously decreasing over the range $\left[-1, 1\right]$. For $\cos\left(\theta_{\mu^{-}}\right)$ distribution, these feature is completely opposite. 
For a quantitative measurement, we define two asymmetry variables $\mathcal{A}_{\pm}$ whose explicit form is written in Eq.~(\ref{Eq:asymmetry}).
\begin{equation}
\mathcal{A}^{\text{NP}}_{\pm} = \frac{\frac{d\sigma}{d\cos\left(\theta_{\mu^{\pm}}\right)}\mid_{\text{range}:= [-1, 0 )} - \frac{d\sigma}{d\cos\left(\theta_{\mu^{\pm}}\right)}\mid_{\text{range} := [0, 1]}}{\frac{d\sigma}{d\cos\left(\theta_{\mu^{\pm}}\right)}\mid_{\text{range} :=[-1, 1]}}
\label{Eq:asymmetry}
\end{equation}
This variable can be effectively used to distinguish between the two scenarios, as it will yield numerical values that differ in their absolute sign, providing a clear and robust means of discrimination.
 However, due to the presence of backgrounds, the distribution of $\cos\left(\theta_{\mu^{\pm}}\right)$ gets distorted. We, therefore, show the normalized distributions of the variables in Figure~\ref{Fig:H0vsHppWbkg}, including the backgrounds. As expected, the variable $\cos\left(\theta_{\mu^{\pm}}\right)$ is skewed towards $\mp 1$ due to the presence of background. However, the characteristic feature, although a bit diminished, is present in the plots. As a result, the asymmetry parameter defined in Eq.~(\ref{Eq:asymmetry}) becomes less sensitive if we take the full range $[-1,1]$. The effectiveness of the parameter is enhanced if the range is changed to $[-0.8,0.8]$. Thus, in the presence of the backgrounds, we define the modified asymmetry parameter as
\begin{eqnarray}
\mathcal{A}^{\text{NP+bkg}}_{\pm} = \frac{\frac{d\sigma}{d\cos\left(\theta_{\mu^{\pm}}\right)}\mid_{\text{range}:= [-0.8, 0 )} - \frac{d\sigma}{d\cos\left(\theta_{\mu^{\pm}}\right)}\mid_{\text{range} := [0, 0.8]}}{\frac{d\sigma}{d\cos\left(\theta_{\mu^{\pm}}\right)}\mid_{\text{range} :=[-0.8, 0.8]}}
\end{eqnarray}

In Table \ref{Tab:asymmetry}, we present the measure of the asymmetry parameters $\mathcal{A}^{\rm NP}_{\pm}$ and $\mathcal{A}^{\rm NP+bkg}_{\pm}$ for the models $\mathcal{L}_{H^{\pm\pm}}$ and $\mathcal{L}_{H^{0}}$, respectively. As argued earlier, the sign of the measure plays the role of a discriminant between the two scenarios. For example, in the case of a doubly charged scalar $H^{\pm\pm}$, the $\mathcal{A}_{+}$ takes a negative value as opposed to the neutral Higgs scenario.   
\begin{table}[ht!]
\centering
\begin{tabular}{|| c | c | c | c ||}
\hline
$\tt{Model~Type}$& $\tt{Benchmark}$ & $\mathcal{A}^{\rm NP}_{\pm}$ & $\mathcal{A}^{\rm NP+bkg}_{\pm}$   \\
\hline
 $\mathcal{L}_{H^{\pm\pm}} \to$~~Eq.(\ref{Eqn:theory})&$m_{H^{\pm\pm}} = \tt{1250~GeV}~\&~h_{\mu\mu} =1$  & $\mp\tt{0.34}$ & $\mp\tt{0.30}$ \\
\hline
$\mathcal{L}_{H^{0}} \to$~~Eq.(\ref{Eq:ModelH0}) & $m_{H^{0}} = \tt{1250~GeV}~\&~\tilde{h}_{\mu\mu} =1$ & $\pm \tt{0.12}$ & $\pm \tt 0.09$  \\
\hline
\end{tabular}
\caption{The measure of asymmetry parameters $\mathcal{A}_{\pm}^{\rm NP}$ and $\mathcal{A}_{\pm}^{\rm NP+bkg}$ are listed for the two different models $\mathcal{L}_{H^{\pm\pm}}$ and $\mathcal{L}_{H^{0}}$, respectively.}
\label{Tab:asymmetry}
\end{table}

\begin{figure}[h]
\includegraphics[width=0.48\textwidth]{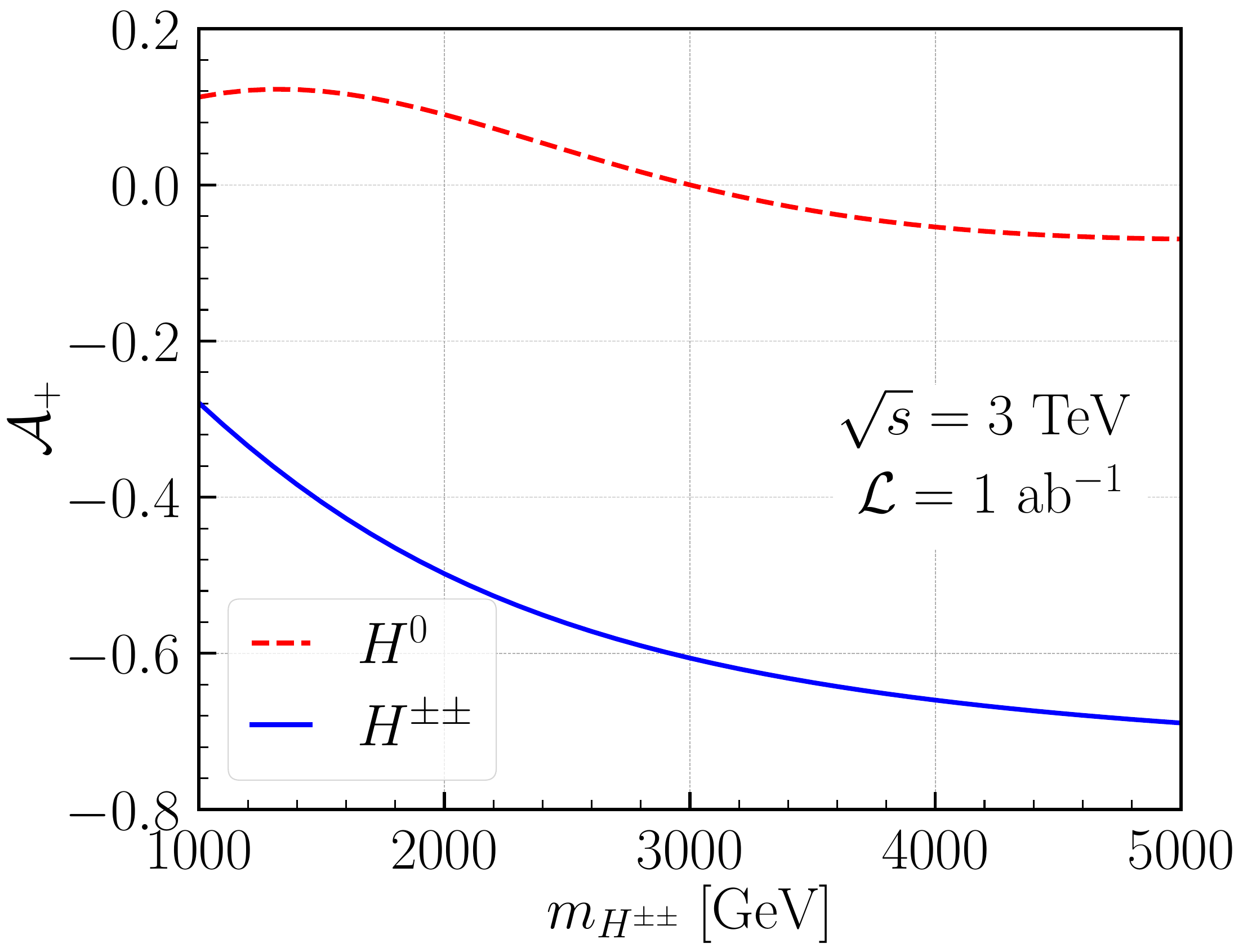}\hfill
\includegraphics[width=0.48\textwidth]{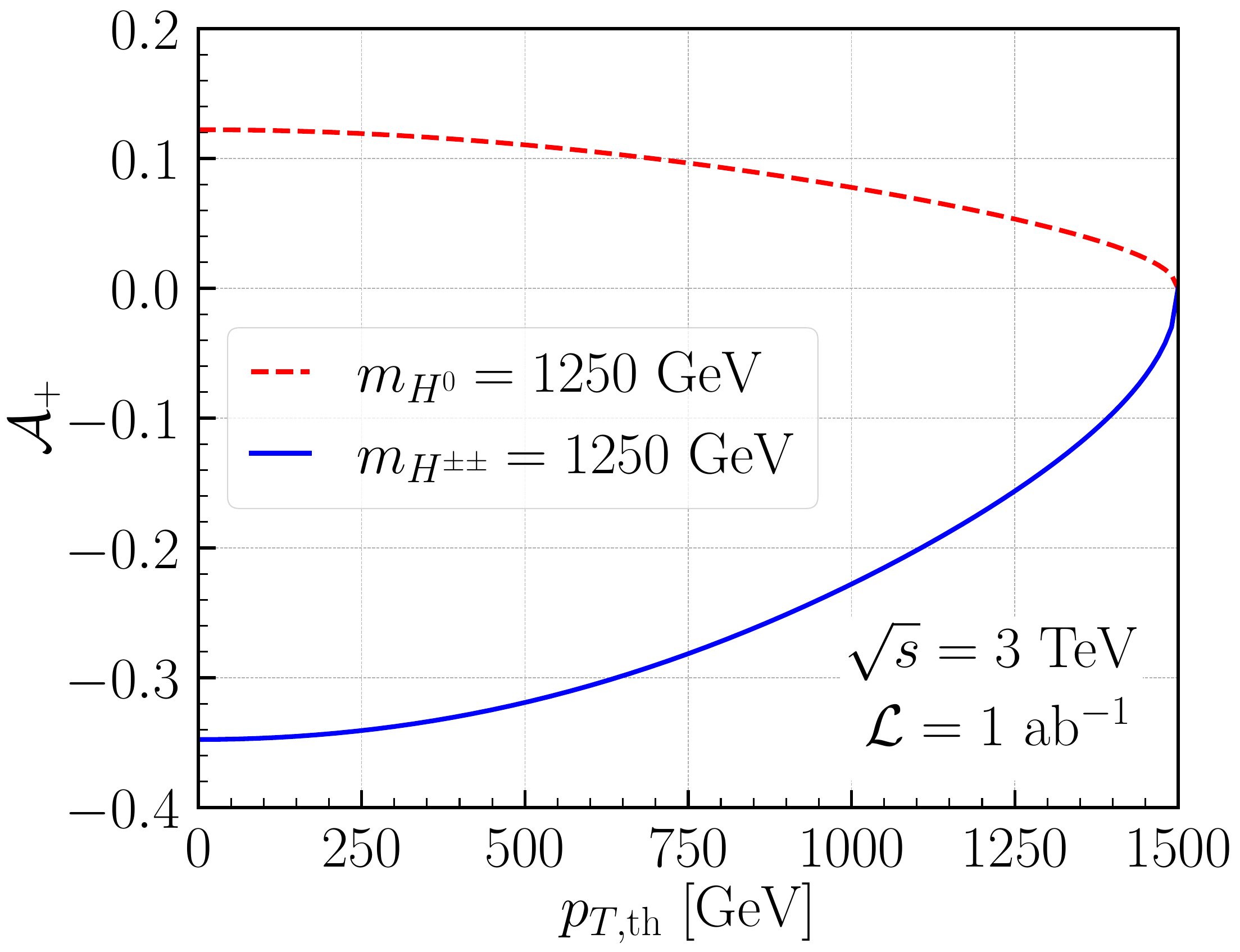}
\caption{Variation of asymmetry parameter as a function of (left) scalar mass and (right) $p_{T}$ threshold on $\mu^\pm$.}
\label{fig:asym_variation}
\end{figure}

There are a few important aspects of the angular variable $\cos\left(\theta_{\mu^\pm}\right)$ and the asymmetry parameter $\mathcal{A}_\pm$ that we would like to highlight. Firstly, neither the angular variable $\cos\left(\theta_{\mu^\pm}\right)$ nor the asymmetry parameter $\mathcal{A}_\pm$ is sensitive to the Yukawa couplings $h_{\mu\mu}$ or $\tilde h_{\mu\mu}$ for the signal. These couplings only appear as an overall scale factor in the distribution, and for the asymmetry parameter, this factor cancels out in the ratio. Furthermore, the asymmetry parameter depends on the mass of the heavy scalar. This dependence is shown in the left panel of Fig.~\ref{fig:asym_variation} for a $\sqrt{s} = 3$~TeV muon collider. In the case of $H^{\pm\pm}$, the asymmetry parameter becomes increasingly negative due to the strong asymmetry in the high-mass region (see Eq.~(\ref{Eq:signalxsec})).

Another important point concerns the effect of selection cuts such as those listed in Table~\ref{Tab:BPmumu1250}. Since our signal topology has no source of missing energy at the hard process -- the only missing energy arises from detector effects -- the $\cos\left(\theta_{\mu^\pm}\right)$ distribution is expected to be affected uniformly. Consequently, the asymmetry parameter should remain almost stable against reasonable cuts on missing transverse momentum. Regarding the transverse momentum of the muons, a cut of $p_T(\mu) > p_{T,\mathrm{th}}$ at a lepton collider, where the center-of-mass energy of the collision equals that of the machine, imposes a restriction on $|\cos\left(\theta_{\mu^\pm}\right)| < \sqrt{1 - \frac{4p_{T,\mathrm{th}}^2}{s}}$ for the two-body final state. Therefore, the $\cos\left(\theta_{\mu^\pm}\right)$ distribution remains unchanged except for this kinematic restriction. However, this restriction modifies the value of $\mathcal{A}_\pm$. This variation is shown in the right panel of Fig.~\ref{fig:asym_variation}. Since the dominant contribution to the asymmetry parameter originates from the region $|\cos\left(\theta_{\mu^\pm}\right)| \to 1$, which is now removed, the value of $\mathcal{A}_\pm$ moves closer to zero and eventually vanishes at $p_{T,\mathrm{th}} = \frac{\sqrt{s}}{2}$.

Finally, the same analysis performed for the $\mu^+\mu^-$ final state can be extended to $e^+e^-$ and $\tau^+\tau^-$ final states. As long as the process is dominated by the $t$-channel and the $e^\pm$ and $\tau^\pm$ are considered massless, the distributions $\cos\left(\theta_{e^\pm}\right)$ and $\cos\left(\theta_{\tau^\pm}\right)$ will be identical to $\cos\left(\theta_{\mu^\pm}\right)$, and hence the asymmetry parameter $\mathcal{A}_\pm$ will take the same value for the $e^+e^-$ and $\tau^+\tau^-$ final states.

We have therefore identified and proposed a crucial variable that could play a pivotal role in distinguishing the same spin particles of different electric charge through an asymmetry. This can also be used for other spin particles mediating similar final states and could be useful at future lepton colliders, including the ILC and CLIC.

\section{Summary and Conclusion}\label{sec:conc}
In this work, we probe the strength of leptonic couplings of a doubly charged scalar at the future muon collider. These exotic scalars naturally appear in different BSM scenarios where the SM scalar spectrum is extended with at least one $\tt{Y = 2}$ non-trivial multiplet. In general, these models can accommodate non-zero neutrino masses, and as a consequence, the doubly charged scalars can couple to SM charged leptons with sizable strength. However, rather than restricting ourselves to a specific $\tt{NP}$ scenario, we adopt a model \emph{agnostic} approach by extending the SM Lagrangian with relevant interaction terms. Much like the $\tt{LEP}$ collider, the proposed muon collider can probe these leptonic couplings in an absolute sense, {\tt i.e.} it can individually constrain couplings such as $\left|h_{\mu e}\right|$, $\left|h_{\mu\mu}\right|$, and $\left|h_{\mu\tau}\right|$ by analyzing $ee$, $\mu\mu$, and $\tau\tau$ final states, respectively.
Our analysis is performed for a muon collider operating at a center-of-mass energy of 3 TeV with an integrated luminosity of $1~\tt{ab^{-1}}$. By carefully analyzing the signal and the relevant $\tt{SM}$ background, we devise a suitable set of rectangular cuts that can suppress the dominant $\tt{SM}$ background while minimally affecting the signal strength. 
Using these optimized cuts, we evaluate the signal significance for doubly charged scalar masses in the range of $\tt{1~TeV}$ to $\tt{10~TeV}$. Using the obtained signal significances, we further extract the discovery reach, presented in the plane of coupling versus mass, where we show that the muon collider can probe a significantly large region of parameter space compared to the existing bounds from flavor experiments and collider searches. The proposed limit on the absolute value of $\left|h_{\mu\mu}\right|$, and $\left|h_{\mu\tau}\right|$ would enrich the existing flavor bounds. 

We also note that similar signals can arise from the $t$-channel exchange of a neutral scalar. However, by studying the angular distribution of the final-state leptons with respect to the muon beam axis, it is possible to discriminate between the doubly charged scalar and the neutral scalar hypotheses. In particular, we find that one can construct a forward-backward (FB) asymmetry variable that exhibits characteristic features: a positive asymmetry score indicating a doubly charged scalar, while a negative score hints towards a neutral scalar.

Thus, a muon collider can offer an optimal environment not only to probe specific second-generation LFV couplings of doubly charged scalars but also to distinguish their effects from those of neutral scalar exchange processes. We would like to reiterate that the primary reason this collider can probe the mass scale of $\tt{NP}$ beyond that of the LHC, and which is much above the $\tt{c.o.m}$ energy of that collider, is because the doubly charged scalar can be produced via a $t$-channel process. This feature of the production mechanism can also be extended to other new physics scenarios, such as $Z'$, leptoquarks, and flavor-violating spin-2 particles. These possibilities provide additional motivation for the construction of a muon collider, which could significantly enhance the scope of future BSM investigations.      
\section*{Acknowledgments}
The work of N.~G.~was supported by the Japan Society for the Promotion of Science (JSPS) as a part of the JSPS Postdoctoral Program (Standard),
grant number: JP24KF0189, and by the World Premier International Research Center Initiative (WPI),
MEXT, Japan (Kavli IPMU). N.~G.~would also like to thank the Centre for High Energy Physics (CHEP), Indian Institute of Science (IISc), for financial support, where part of the work was done. A.~S.~thanks Anusandhan
National Research Foundation (ANRF) for providing the necessary financial support through the SERB-NPDF grant
(Ref No: PDF/2023/002572). S.~K.~R.~acknowledges the support from the Department of Atomic Energy (DAE), India, for the Regional Centre for Accelerator-based Particle Physics (RECAPP), Harish Chandra Research Institute. A.~S.~would also like to thank Dr. Siddharth P. Maharathy for providing useful suggestions on the event simulation portion of the work.

\section*{Data Availability Statement}
This study is based on simulated data generated using standard HEP packages. The details of the simulation process that support our findings are provided within the article.

\appendix
\section{Interference between the SM $\gamma/Z$ and $H^{\pm\pm}$}
\label{sec:appA}

\begin{figure}[h]
\begin{center}
\includegraphics[width=0.5\textwidth]{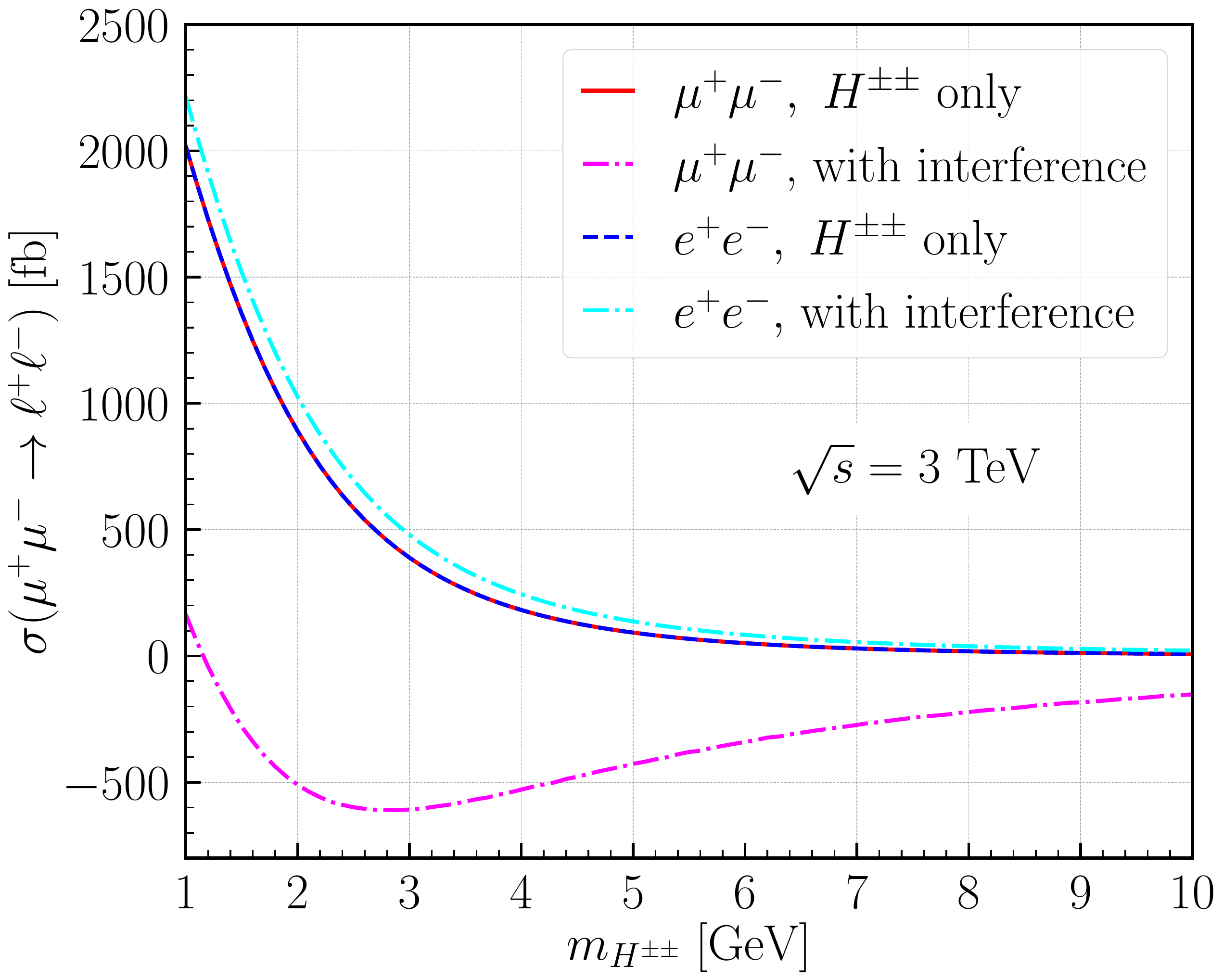}
\end{center}
\caption{Figure demonstrating the interference effects in total cross section in the $e^+e^-$ and $\mu^+\mu^-$ channels.}
\label{fig:crossxinterference}
\end{figure}

In this section, we discuss the interference effect of the SM $\gamma/Z$ exchange diagrams with the $H^{\pm\pm}$ exchange diagrams. 
In the $\mu^+\mu^- \to e^+ e^-$ process, the interference effect arises from the $H^{\pm\pm}$ exchange diagram (Figure~\ref{Fig:feynman}) is the $t$-channel and $\gamma/Z$ exchange diagrams in the $s$-channel. The contribution from the interference effect is positive, and it is approximately 10\%. Importantly, the interference effect does not change the shape of the distribution of $E^{\text{Miss}}_{T}$ and $p_T (\ell_1)$ as given in Figure~\ref{Fig:eehist}. In Figure~\ref{fig:crossxinterference}, we show the variation of new physics contribution to the cross section of $\mu^+\mu^- \to e^+ e^-$ process at 3~TeV muon collider as a function of $m_{H^{\pm\pm}}$. In the figure, the blue dashed line represents the contribution from the $H^{\pm\pm}$ exchange diagram only, and the cross-section, including the interference effect, is represented by the cyan dash-dotted line.

For $\mu^+\mu^- \to \mu^+ \mu^-$ process, the contribution of interference comes from the $t$-channel $H^{\pm\pm}$ exchange diagram (Figure~\ref{Fig:feynman}) with $s$- and $t$-channel $\gamma/Z$ exchange diagrams. In this case, due to the presence of $t$-channel $\gamma/Z$ exchange diagrams, the effects contribute negatively, and the amount is also large. These effects are demonstrated in the plots of cross-section due to new physics in Figure~\ref{fig:crossxinterference}. In the figure, red solid lines represent the cross-section for $H^{\pm\pm}$ exchange diagrams only, and the magenta dashed-dotted line represents the contribution to the contribution when interference effects are considered. The interference effect is highly negative, making the total additional contribution from the BSM negative. 

\begin{figure}[h]
\includegraphics[width=0.48\textwidth]{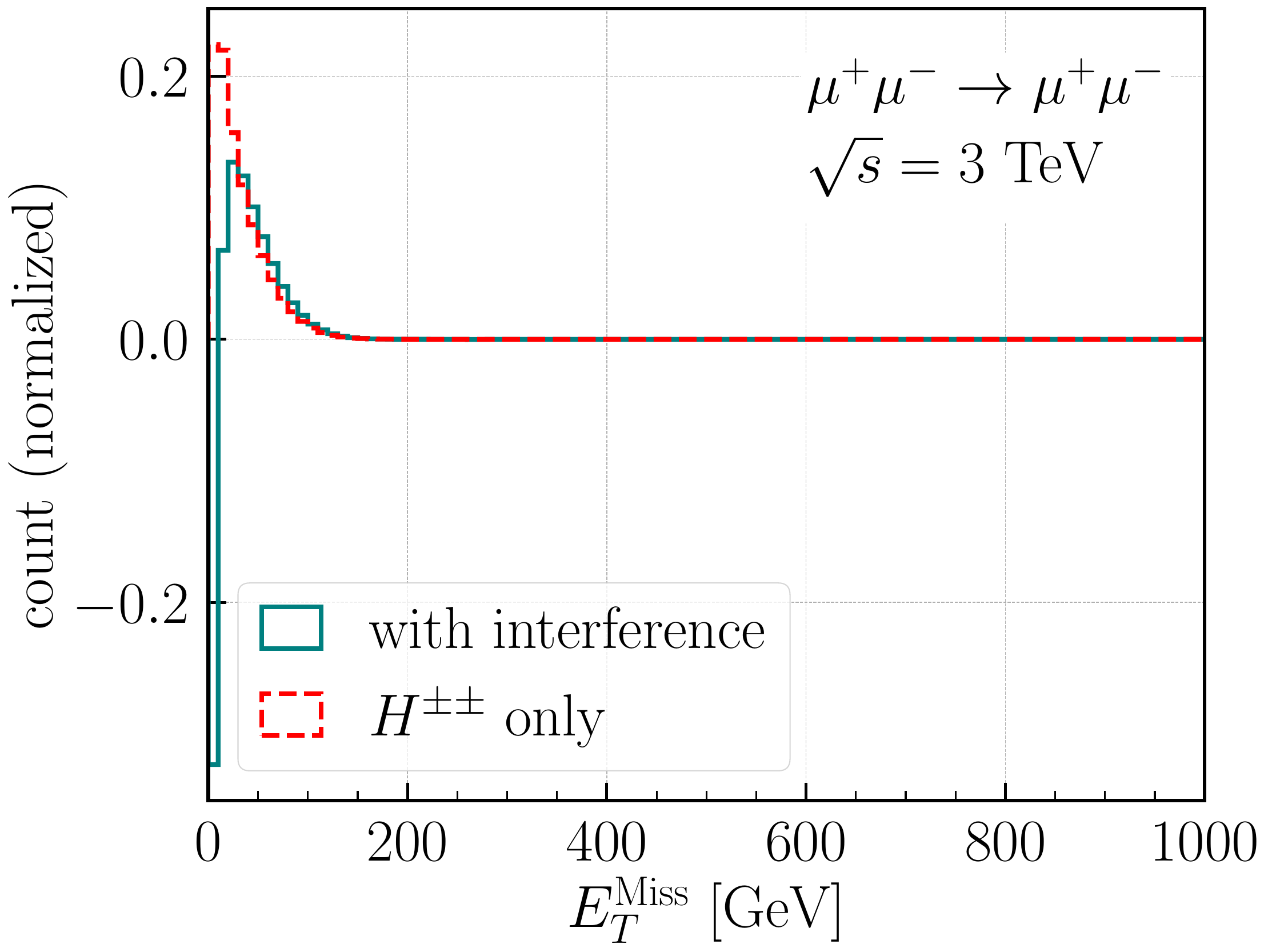}
\includegraphics[width=0.48\textwidth]{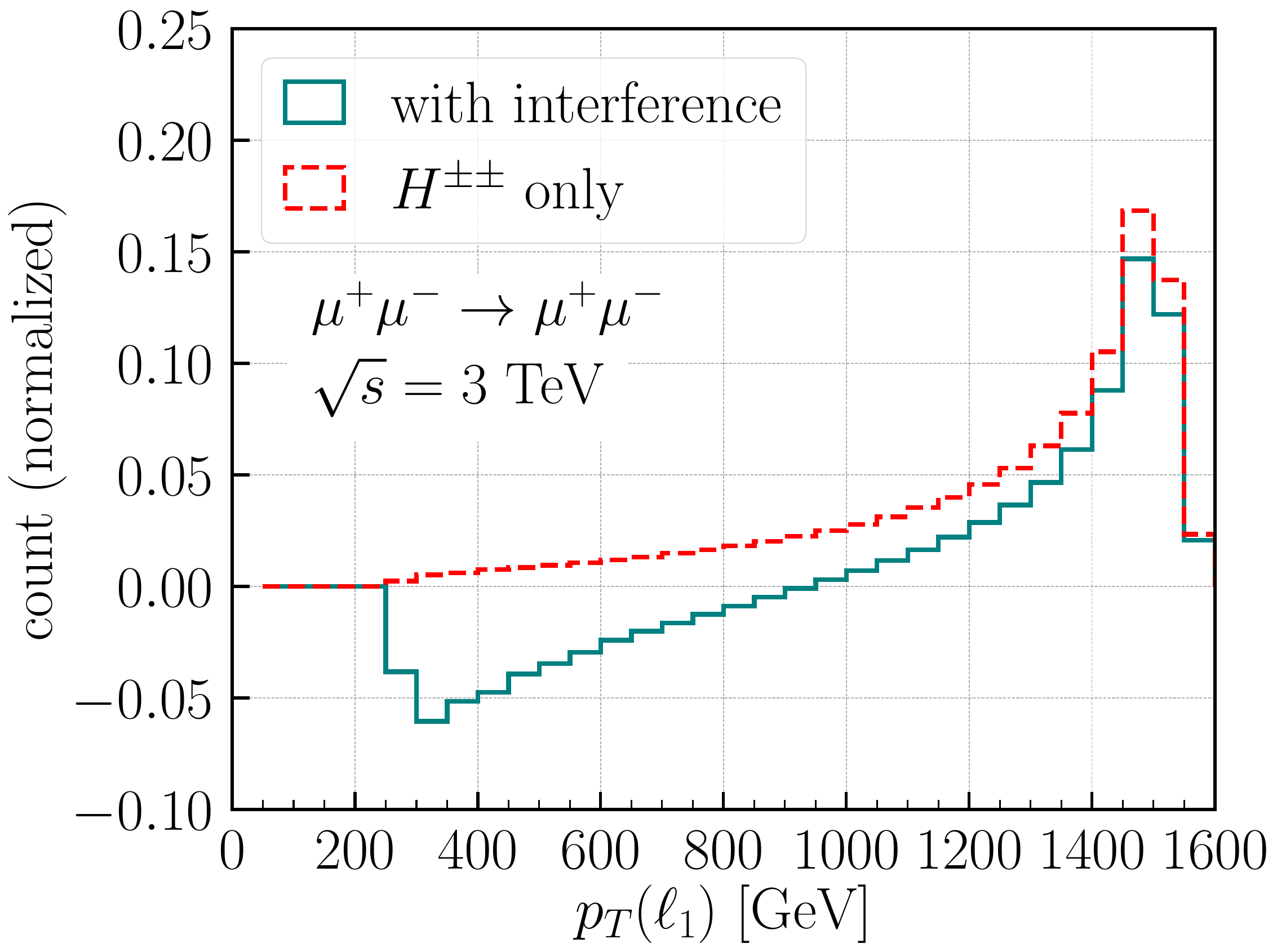}
\caption{Distributions of $E^{\rm Miss}_T$ and $p_T(\ell_1)$ in $\mu\mu$ channel with and without interference effects.}
\label{fig:mumu-dist}
\end{figure}

In Figure~\ref{fig:mumu-dist}, the distributions of $E^{\rm Miss}_T$ and $p_T(\ell_1)$ in $\mu\mu$ channels. In all the figures, there are two plots corresponding to contributions due to only the $H^{\pm\pm}$ exchange diagram (red dashed) and contributions including interference effects (teal solid). The first bin in $E^{\rm Miss}_T$ histogram and bins up to 950~GeV in $p_T(\ell_1)$ histogram are negative, indicating that the interference part is highly negative in that part of the phase space region.

\begin{table}[h]
\begin{center}
\begin{tabular}{|c|c|c|c|}
\hline
Cut & \tt $\sigma_{H^{++}\text{-only}}$ [fb] & \tt $\sigma_{\text{SM}+H^{++}}-\sigma_\text{SM-only} \equiv \sigma_\text{eff}$ [fb] & \tt $\dfrac{\sigma_\text{eff}-\sigma_{H^{++}\text{-only}}}{\sigma_{H^{++}\text{-only}}}$ \\
\hline
\hline
${\tt N}_e = 2$  & $\tt 1.151\times10^3$ & $\tt 1.274\times10^3$ & \tt 10.74\% \\
\hline
\tt $E^{\tt Miss}_T < \tt 50$ GeV & $\tt 1.137\times10^3$ & $\tt 1.259\times10^3$ & \tt 10.75\% \\
\hline
\tt ${p_T}(e_1) > \tt 1300$ GeV & $\tt 6.540\times10^2$ & $\tt 7.311\times10^2$ & \tt 9.65\% \\
\hline
\end{tabular}
\caption{Effect of interference in cut-flow as defined in Table~\ref{Tab:BPee1250} for $ee$ final state at $\sqrt{s}=3$ TeV and $m_{H^{\pm\pm}}=1250$ GeV.}\label{tab:ee}
\end{center}
\vspace{-10pt}
\end{table}

In Tables~\ref{tab:ee}--\ref{tab:tata}, we show the cross sections at different cuts applied form $m_{H^{\pm\pm}}$. The $\sigma_{H^{\pm\pm}\rm -only}$ column is when only the $H^{\pm\pm}$ exchange diagram is considered. The column to the right of that contains the cross-section when the interference terms are included. The last column shows the relative change with respect to the $H^{\pm\pm}$-only contribution. In the case of the $ee$ and $\tau\tau$ final states, the interference effects contribute to approximately an additional 10\%. The interference effects are negative in the case of the $\mu\mu$ channel.

\begin{table}[h]
\begin{center}
\begin{tabular}{|c|c|c|c|c|c|}
\hline
Cut & \tt $\sigma_{H^{++}\text{-only}}$ [fb] & \tt $\sigma_{\text{SM}+H^{++}}-\sigma_\text{SM-only}\equiv \sigma_\text{eff}$ [fb] & \tt$\dfrac{\sigma_\text{eff}-\sigma_{H^{++}\text{-only}}}{\sigma_{H^{++}\text{-only}}}$  \\
\hline
\hline
$\tt N_\mu = 2$ & $\tt 1.574\times10^3$ & $\tt 3.585\times10^2$ & $\tt -77.23\%$ \\
\hline
\tt $\tt E^{\rm Miss}_T < 100$ GeV & $\tt 1.542\times10^3$ & $\tt 3.303\times10^2$ & $\tt -78.57\%$ \\
\hline
\tt $\tt {p_T}(\mu_1) > 1350$ GeV & $\tt 7.840\times10^2$ & $\tt 6.870\times10^2$ & $\tt -12.38\%$ \\
\hline
\end{tabular}
\caption{Effect of interference in cut-flow as defined in Table~\ref{Tab:BPmumu1250} for $\mu\mu$ final state at $\sqrt{s}=3$ TeV and $m_{H^{\pm\pm}}=1250$ GeV.}\label{tab:mumu}
\end{center}
\end{table}

\begin{table}[h!]
\centering
\begin{tabular}{|c|c|c|c|}
\hline
Cut & \tt $\sigma_{H^{++}\text{-only}}$ [fb] & \tt $\sigma_{\text{SM}+H^{++}}-\sigma_\text{SM-only}\equiv \sigma_\text{eff}$ [fb] & \tt$\dfrac{\sigma_\text{eff}-\sigma_{H^{++}\text{-only}}}{\sigma_{H^{++}\text{-only}}}$  \\
\hline
\hline
$\tt{N_{e} = 1~\&~N_{\mu} = 1}$  &  $\tt{7.36\times10^{1}}$ & $\tt{8.17\times10^{1}}$ & \tt{11.06\%}  \\
\hline
\tt{$\left|\cos\theta_{\mu}\right| <$ 0.8} &  $\tt{6.63\times10^{1}}$ &  $\tt{7.34\times10^{1}}$ & $\tt{10.65\%}$ \\
\hline
\end{tabular}\caption{Effect of interference in cut-flow as defined in Table~\ref{Tab:BPtata1250} for $\tau\tau$ final state at $\sqrt{s}=3$ TeV and $m_{H^{\pm\pm}}=1250$ GeV.}\label{tab:tata}
\end{table}

From the above explanations, we can see that the interference effect changes the cross-section before the final cut is applied. However, after the final selection cut, the change is only about 10\%. The expectation in the $\tau\tau$ final state is very similar to the $ee$ final state since the SM $\gamma/Z$ exchange diagram is only the $s$-channel diagram in both cases. The corresponding distributions for $\tau\tau$ final state shown in Figures~\ref{Fig:tatahistpTeT} and \ref{Fig:tatahist} do not change their shapes due to the incorporation of the interference effects.


\providecommand{\href}[2]{#2}\begingroup\raggedright\endgroup

\end{document}